\documentclass[aps,prd,longbibliography,singlecolumn,superscriptaddress,preprintnumbers,floatfix,nofootinbib,notitlepage,showkeys]{revtex4-1}
\usepackage[utf8]{inputenc}
\usepackage{graphicx}
\usepackage{tabularx}
\usepackage{hyphenat}
\usepackage{amsmath,amssymb,amsfonts,amsthm,latexsym}
\usepackage{bbold}
\usepackage{slashed}
\usepackage[table,xcdraw,svgnames,dvipsnames]{xcolor}
\usepackage{siunitx}
\usepackage{caption}
\usepackage{subcaption}
\usepackage{lineno}
\usepackage[colorlinks]{hyperref}
\AtBeginDocument{\hypersetup{citecolor=OliveGreen,linkcolor=BrickRed,urlcolor=MidnightBlue}}
\usepackage{mathtools}
\usepackage{placeins}

\newcommand{\bra}{\left\langle}
\newcommand{\ket}{\right\rangle}
\mathchardef\mhyphen="2D 

\usepackage{graphicx}
\begin{document}

\title{Static force from generalized Wilson loops on the lattice using the gradient flow}

\author{Nora Brambilla}
\email{nora.brambilla@tum.de}
\affiliation{Technical University of Munich, TUM School of Natural Sciences, Physics Department, James-Franck-Strasse 1, 85748 Garching, Germany}
\affiliation{Institute for Advanced Study, Technical University of Munich,
Lichtenbergstrasse 2a, 85748 Garching, Germany}
\affiliation{Munich Data Science Institute, Technical University of Munich, \\
Walther-von-Dyck-Strasse 10, 85748 Garching, Germany}

\author{Viljami Leino}
\email{viljami.leino@uni-mainz.de}
\affiliation{Helmholtz Institut Mainz, Staudingerweg 18, 55128 Mainz, Germany}
\affiliation{Institut für Kernphysik, Johannes Gutenberg-Universität Mainz, 
             Johann-Joachim-Becher-Weg 48, 55128 Mainz, Germany}

\author{Julian Mayer-Steudte}
\email{julian.mayer-steudte@tum.de}
\affiliation{Technical University of Munich, TUM School of Natural Sciences, Physics Department, James-Franck-Strasse 1, 85748 Garching, Germany}
\affiliation{Munich Data Science Institute, Technical University of Munich, \\
Walther-von-Dyck-Strasse 10, 85748 Garching, Germany}

\author{Antonio Vairo}
\email{antonio.vairo@tum.de}
\affiliation{Technical University of Munich, TUM School of Natural Sciences, Physics Department, James-Franck-Strasse 1, 85748 Garching, Germany}
\date{Received 26 January 2024; accepted 23 May 2024} 
\preprint{TUM-EFT 175/22, MITP-23-080} 

\begin{abstract}
    The static QCD force from the lattice can be used to extract $\Lambda_{\overline{\textrm{MS}}}$, which determines the running of the strong coupling. Usually, this is done with a numerical derivative of the static potential. However, this introduces additional systematic uncertainties; thus, we use another observable to measure the static force directly. This observable consists of a Wilson loop with a chromoelectric field insertion. We work in the pure SU(3) gauge theory. We use gradient flow to improve the signal-to-noise ratio and to address the field insertion. We extract $\Lambda_{\overline{\textrm{MS}}}^{n_f=0}$ from the data by exploring different methods to perform the zero-flow-time limit. We obtain the value $\sqrt{8t_0} \Lambda_{\overline{\textrm{MS}}}^{n_f=0} =0.629^{+22}_{-26}$, where $t_0$ is a flow-time reference scale. We also obtain precise determinations of several scales: $r_0/r_1$, $\sqrt{8 t_0}/r_0$, $\sqrt{8 t_0}/r_1$ and we compare these to the literature. 
    The gradient flow appears to be a  promising method for calculations of  Wilson loops 
    with chromoelectric and chromomagnetic insertions in quenched and unquenched configurations.\\

    \noindent DOI: \href{https://doi.org/10.1103/PhysRevD.109.114517}{10.1103/PhysRevD.109.114517}
\end{abstract}

\maketitle
\vspace{1cm}
\section{Introduction}
The Standard Model of particle physics is one of the most precisely tested theories.
A precise knowledge of the Standard Model parameters is a necessary condition to work out accurate perturbative predictions, and to compare them with high-precision experimental measurements. Quantum chromodynamics (QCD) is 
the sector of the Standard Model that describes the strong interaction. It is a field theory based on the gauge group SU(3) that depends on just one coupling, $g$, or equivalently $\alpha_s = g^2/(4\pi)$. 
The coupling may be traded at any time with an intrinsic scale; in the $\overline{\textrm{MS}}$ scheme, this is  $\Lambda_{\overline{\textrm{MS}}}$.
Once renormalized, $\alpha_s$ is small at energy scales much larger than $\Lambda_{\overline{\textrm{MS}}}$, 
a property known as asymptotic freedom, but it becomes of order 1 at energy scales close to 
$\Lambda_{\overline{\textrm{MS}}}$. At high energies, we can rely on weak coupling perturbation 
theory to compute QCD observables.

The value of $\Lambda_{\overline{\textrm{MS}}}$, or equivalently $\alpha_s$, at a large energy scale can be determined by comparing some high-energy observable computed in weak coupling perturbation theory with data.
A viable alternative is to replace data with lattice QCD computations i.e., the exact evaluation 
of the observable in QCD via Monte Carlo computations.
For the current status of $\Lambda_{\overline{\textrm{MS}}}$ extractions from lattice QCD, see for example the recent reviews~\cite{FlavourLatticeAveragingGroupFLAG:2021npn,dEnterria:2022hzv}. While in the last several years lattice extractions of $\Lambda_{\overline{\textrm{MS}}}$ have been mostly done in QCD with dynamical quarks, the interest towards the running of the coupling in the pure gauge version of QCD i.e., without dynamical quarks, also called quenched QCD has been recently reignited~\cite{DallaBrida:2022eua}.
With modern lattice methods, the extraction of the coupling from the pure gauge theory can be done much more precisely nowadays than in the past, when quenched calculations were the only viable option.

The static energy is a well-understood quantity in lattice QCD~\cite{Necco:2001xg,Guagnelli:1998ud} that can be used to set the lattice scale~\cite{Sommer:1993ce}. Furthermore, the static energy is an observable  that can be used to extract $\Lambda_{\overline{\textrm{MS}}}$ by comparing its perturbative expression with lattice data at short distances.
The QCD scale $\Lambda_{\overline{\textrm{MS}}}$ has been extracted from the static energy both in pure gauge~\cite{Bali:1992ru,Booth:1992bm,Brambilla:2010pp,Husung:2017qjz,Husung:2019ytz} and with dynamical fermions~\cite{Jansen:2011vv,Karbstein:2014bsa,Bazavov:2012ka,Bazavov:2014soa,Takaura:2018vcy,Bazavov:2019qoo,Ayala:2020odx}.

In lattice QCD, the static energy suffers from a linear divergence and needs to be renormalized.
In dimensionally regularized perturbation theory, 
this divergence becomes a renormalon of mass dimension 1.
For these reasons, it presents some advantages to look at the derivative of the static energy, which is the static force. The static force is not affected by linear divergences in lattice QCD or by renormalons of mass dimension 1 in dimensional regularization, yet it contains all the relevant 
information on the running of the strong coupling, as this is entirely encoded in the slope of the static energy.
The force is known up to next-to-next-to-next-to-leading logarithmic accuracy in the coupling~\cite{Brambilla:1999qa,Pineda:2000gza,Brambilla:2006wp,Brambilla:2009bi,Anzai:2009tm,Smirnov:2009fh}. 

The derivative of the static energy performed numerically on lattice data introduces additional systematic uncertainties. Therefore, it may be advantageous to compute the force directly from a suitable observable~\cite{Vairo:2015vgb,Vairo:2016pxb,Brambilla:2000gk}. This observable consists of a Wilson loop with a chromoelectric field insertion. A difficulty related to this observable is that field insertions on Wilson loops when evaluated on the lattice have a bad signal-to-noise ratio and a slow convergence to the continuum limit originating from the discretization of the field components. This was studied in a previous work~\cite{Brambilla:2021wqs}. 
In this work, to overcome the difficulty, we rely on the gradient flow method~\cite{Narayanan:2006rf, Luscher:2009eq, Luscher:2010iy} to improve the signal-to-noise ratio and to remove the discretization effects of the field components.
Wilson loops smeared with gradient flow have been studied before in terms of Creutz ratios~\cite{Risch:2022vof,Risch:2023qjy,Okawa:2014kgi}. In Ref.~\cite{Bazavov:2023dci}, the Wilson line correlator in the Coulomb gauge was measured at finite $T$ with gradient flow.  
To our knowledge, this was the first time the gradient flow was applied to the force directly.
Furthermore, this study also serves as a preparation for the study of similar Wilson loops with field insertions appearing in the computation of several observables in the context of nonrelativistic effective field theories. New methods for integrating the gradient flow equations, based on Runge-Kutta methods, have been developed and implemented during the last few years~\cite{Bazavov:2021pik}. The force in gradient flow, besides that with lattice QCD, can also be computed analytically in perturbation theory.
In perturbation theory, the static force at finite flow time is known at one-loop order~\cite{Brambilla:2021egm}.

The paper is structured as follows: In Sec.~\ref{sec:theoretical_background}, we discuss the theoretical background: we introduce the gradient flow and the static force at zero and finite flow time, in continuum and on the lattice. The lattice setup is described in Sec.~\ref{sec:lattice_setup}, and in Sec.~\ref{sec:prep_analysis} we show our numerical results and perform the continuum limits. Finally, in Sec.~\ref{sec:continuum_analysis}, we discuss our results and extract $\Lambda_{\overline{\textrm{MS}}}^{n_f=0}$.
Preliminary results based on these data have appeared before in conference proceedings~\cite{Leino:2021vop,Mayer-Steudte:2022uih}.

\section{Theoretical background}\label{sec:theoretical_background}
\subsection{The static force}\label{sec:static_force}
The static energy $V(r)$ in Euclidean QCD
is related to a rectangular Wilson loop $W_{r\times T}$ with temporal extent from $0$ to $T$ and spatial extent  $r$~\cite{Wilson:1974sk} by
\begin{linenomath}\begin{align}
    V(r) &= -\lim_{T\rightarrow \infty} \frac{\ln\langle \mathrm{Tr}(P\,W_{r\times T})\rangle}{T}\nonumber \\
    &= - \frac{1}{a}\lim_{T\rightarrow \infty} \frac{\langle \mathrm{Tr}(P\,W_{r\times(T+a)})\rangle}{\langle \mathrm{Tr}(P\,W_{r\times T})\rangle}\,,\label{eq:static_energy_definition}\\
    W_{r\times T} &=  \left\{ \exp\left( i\oint _{r\times T} dz_\mu gA_\mu \right) \right\}\,,
    \label{eq:Wilsonlinedef}
\end{align}\end{linenomath}
where $a$ is the lattice spacing, $g$ the strong coupling, Tr the trace over the color matrices,  and $P$ is the path-ordering operator for the color matrices. 
In dimensional regularization, the static energy has a renormalon ambiguity of order $\Lambda_\mathrm{QCD}$,
while on the lattice there is a linear divergence of order $1/a$ coming from the self-energy of the Wilson
line.
Both the perturbative and lattice problems manifest as a constant shift to the potential. This may be renormalized by fixing the potential to a given value at a given point $r^\ast$:
\begin{linenomath}
    \begin{align}
        V^r(r) = V(r) - V(r^*) \,.
    \end{align}
\end{linenomath}
Alternatively, taking the derivative removes the divergent constant, and we obtain a renormalized quantity, the static force.

The static force $F(r)$ is defined as the derivative of the static energy:
\begin{linenomath}\begin{align}
    F_{\partial V}(r) = \partial_r V(r) \,.
\end{align}\end{linenomath}
In perturbation theory, the static force is known up to next-to-next-to-next-to-leading logarithmic order (N$^3$LL)~\cite{Brambilla:1999qa,Pineda:2000gza,Brambilla:2006wp,Brambilla:2009bi,Anzai:2009tm,Smirnov:2009fh}.
On the lattice, this derivative is evaluated from the static energy data either with interpolations or with
finite differences, which leads to increased systematic errors.
It is possible, however, to carry out the derivative of the Wilson lines at the level of Eq.~\eqref{eq:Wilsonlinedef}, and rewrite the force as~\cite{Vairo:2015vgb,Vairo:2016pxb,Brambilla:2000gk}
\begin{linenomath}\begin{align}
    F_E(r) &= -\lim _{T\rightarrow\infty}\frac{i}{\langle \mathrm{Tr}(P\, W_{r\times T} \rangle} \left\langle\mathrm{Tr}\left( P\left\{ \exp\left( i\oint _{r\times T} dz_\mu gA_\mu \right) \mathbf{\hat{r}}\cdot g\mathbf{E}(\mathbf{r}, t^*) \right\} \right) \right\rangle \\
    &= -\lim _{T\rightarrow\infty} i \frac{\langle \mathrm{Tr}\{ P\,W_{r\times T} \mathbf{\hat{r}}\cdot g\mathbf{E}(\mathbf{r},t^* \} \rangle}{\langle \mathrm{Tr}(P\,W_{r\times T} \rangle}\,,\label{eq:static_force_E_insertion_definition}
\end{align}\end{linenomath}
where the expression in the numerator consists of a static  Wilson loop with a chromoelectric field insertion on the temporal Wilson line at position $t^*$, and $\mathbf{\hat{r}}$ is the spatial direction of the quark-antiquark pair separation; $t^*$ can be chosen arbitrarily. Since both expressions for the force represent the same renormalized quantity, it holds in the continuum that $F_E = F_{\partial V} = F$.

\subsection{Gradient flow}
We rely on the gradient flow method \cite{Narayanan:2006rf, Luscher:2009eq, Luscher:2010iy} for measuring Wilson loops with and without chromoelectric field insertions. The gradient flow is a continuous transformation of the gauge fields toward the minimum of the Yang-Mills gauge action along a fictitious flow time $\tau_F$:
\begin{linenomath}
    \begin{align}
        \Dot{B}_\mu(\tau_F,x) &= D_\nu G_{\nu\mu}=-g_0^2\frac{\delta S_\mathrm{YM}[B]}{\delta B_\mu(\tau_F,x)}\label{eq:gradient_flow_definition}\,,\\
        B_{\mu |\tau_F=0} &= A_\mu \,,\\
        G_{\mu\nu} &= \partial_\mu B_\nu - \partial_\nu B_\mu + [B_\mu,B_\nu ], D_\mu = \partial_\mu + [B_\mu, .]\,,
    \end{align}
\end{linenomath}
where $B_\mu(\tau_F)$ are the flowed gauge fields at flow time $\tau_F$ with the SU(3) QCD gauge fields $A_\mu$ as the initial condition at zero flow time, $S_\mathrm{YM}$ is the Yang-Mills action evaluated with the flowed gauge fields; $G_{\mu\nu}$ is the field strength tensor evaluated with the flowed gauge fields; and $D_\mu$ is the gauge covariant derivative. The flow depends on the local neighboring gauge field values through the derivative of the action with respect to the gauge field at position $x$, and its characteristic range is given by the flow radius $\sqrt{8\tau_F}$. This results in a smearing that 
cools off systematically the ultraviolet physics and automatically renormalizes gauge-invariant observables~\cite{Luscher:2010we, Luscher:2011bx}.
Furthermore,  we introduce the reference scale $t_0$~\cite{Luscher:2010iy}, defined implicitly through the expectation value of the action density
\begin{equation}
    E = \frac{1}{4}G_{\mu\nu}^aG_{\mu\nu}^a\,,
\end{equation}
as
\begin{equation}
    \tau_F^2\langle E\rangle |_{\tau_F=t_0} = 0.3\,.
    \label{eq:t0_definition}
\end{equation}
The gradient flow equation is adapted for flowed link variables $V_{\tau_F}(\mu,x)$ on the lattice as
\begin{align}
    \Dot{V}_{\tau_F}(x,\mu) &= -g_0^2\left(\partial_{x,\mu} S_\mathrm{Gauge}(V_{\tau_F})\right) V_{\tau_F}(x,\mu)\label{eq:gradient_flow_definition_lattice}\,, \\
    V_{\tau_F}(x,\mu)|_{\tau_F=0} &= U_\mu(x)\,,\nonumber
\end{align}
where $S_\mathrm{Gauge}(V_{\tau_F})$ is some lattice gauge action evaluated with the flowed link variables, $\partial_{x,\mu}$ is the derivative with respect to
$V_{\tau_F}(x,\mu)$, 
and $U_\mu(x)$ is the original SU(3) link variable. The flowed link variables of a gauge field configuration depend uniquely on the initial gauge field configuration $U$ -- i.e., $V_{\tau_F}=V_{\tau_F}(U)$ -- and flowed observables are obtained by replacing the original link variables with the flowed link variables: $O(\tau_F)=O|_{U=V_{\tau_F}}$. The flowed expectation value of $O$ is evaluated on the flowed gauge ensemble and can be written as
\begin{align}
    \bra O(\tau_F)\ket &= \frac{1}{Z} \int \mathcal{D}[U]e^{-S_E[U]}O[V_{\tau_F}(U)]\,,
\end{align}
which is still a path integral with the Euclidean action $S_E$ of the original zero-flow-time theory. 
Therefore, on the lattice, the expectation value is given by
\begin{align}
    \bra O(\tau_F)\ket \approx \frac{1}{N}\sum_{U,p(U)\propto e^{-S_E}} O(V_{\tau_F}(U))\,,
\end{align}
where $N$ is the number of gauge fields.
We solve the gradient flow on the lattice by an iterative Runge-Kutta implementation for the SU(3) matrices. We use either a fixed step-size algorithm \cite{Luscher:2010iy} or an adaptive step-size algorithm \cite{Fritzsch:2013je, Bazavov:2021pik}.

\subsection{The perturbative static force at finite flow time}
The one-loop formula for the static force in gradient flow is \cite{Brambilla:2021egm}
\begin{align}
    F(r,\tau_F) = &\frac{\alpha_S(\mu)C_F}{r^2}\Big[\left( 1+ \frac{\alpha_S}{4\pi}a_1 \right) \mathcal{F}_0(r,\tau_F) + \nonumber\\
    &\frac{\alpha_S}{4\pi}\beta_0 \mathcal{F}_\mathrm{NLO}^L(r,\tau_F,\mu) + \frac{\alpha_SC_A}{4\pi}\mathcal{F}_\mathrm{NLO}^F(r,\tau_F) \Big] \nonumber \\
    &+ \mathcal{O}\left(\alpha_S^3\right), \label{eq:flowed_force_full_1_loop_formula}
\end{align}
with $C_F=(N_C^2-1)/(2N_C)$, $C_A=N_C$, $N_C=3$ the number of colors, $\beta_0= {11}C_A/3-2n_f/3$, $n_f$ the number of flavors, and $a_1 = 31C_A/9-10n_f/9$. The functions $\mathcal{F}_0(r,\tau_F)$ and $\mathcal{F}_\mathrm{NLO}^L(r,\tau_F,\mu)$ are given analytically with
\begin{align}\label{eq:f0def}
    \mathcal{F}_0(r,\tau_F) =\ &\mathrm{erf}\left( \frac{r}{\sqrt{8\tau_F}} \right) - \frac{r}{\sqrt{2\pi \tau_F}}\exp \left( -\frac{r^2}{8\tau_F} \right),
\end{align}
\begin{align}
    \mathcal{F}_\mathrm{NLO}^L(r,\tau_F,\mu) = \ &\log (\mu^2r^2)\mathcal{F}_0(r,\tau_F) + \log \left(\frac{8\tau_F}{r^2}e^{\gamma_\mathrm{E}}\right) \mathcal{F}_0(r,\tau_F) \nonumber \\
    &- \frac{r}{\sqrt{2\pi\tau_F}} \bigg[ e^{-\frac{r^2}{8\tau_F}} M^{(1,0,0)}\left( 0,\frac{1}{2},\frac{r^2}{8\tau_F}\right) + M^{(1,0,0)}\left(\frac{1}{2},\frac{3}{2},-\frac{r^2}{8\tau_F}\right) \bigg],
\end{align}
where $\gamma_\mathrm{E}$ is the Euler-Mascheroni constant and $M(a,b,z)$ is the confluent hypergeometric function defined by
\begin{equation}
    M(a,b,z) = \sum _{k=0}^\infty \frac{(a)_k}{(b)_k}\frac{z^k}{k!}\,,
\end{equation}
with $(x)_k=\Gamma(x+k)/\Gamma(x)$, and
\begin{equation}
    M^{(1,0,0)}(a,b,z) = \frac{\partial}{\partial a}M(a,b,z)\,.
\end{equation}
For $r > \sqrt{\tau_F}$, we approximate $\mathcal{F}_\mathrm{NLO}^F$ with the polynomial
\begin{align}
    \mathcal{F}_\mathrm{NLO}^F(\xi=r/\sqrt{\tau_F}) = &\sum_{n=1}^{10} \frac{c_n}{n!}\left( \frac{\xi}{1+\xi/C_a}\right) ^ne^{-\xi} + \frac{44-C_b}{C_a+\xi^2} + \frac{C_b\xi^2}{\xi^4+C_c}\,,
    \label{eq:Ffnlo_polynom}
\end{align}
where $C_a=109.358$, $C_b = 43.8438$, $C_c=404.790$, and $c_n$ values are listed in Table \ref{tab:cn_coeff}. In practice, the computation of $M^{(1,0,0)}$ takes a considerable amount of time, which is a problem for fitting this function. Those terms depend only on the flow-time ratio $\tau_F/r^2$; hence, we precompute them on a fine flow-time ratio grid, and we use spline interpolations for further calls of the perturbative formula.

\begin{table}[]
    \centering
    \begin{tabular}{c|c}
    $c_1$ & -0.0501648 \\
    $c_2$ & 0.526758 \\
    $c_3$ & -5.55177 \\
    $c_4$ & 45.8753 \\
    $c_5$ & -147.8 \\
    $c_6$ & 463.906 \\
    $c_7$ & -851.741 \\
    $c_8$ & 884.315 \\
    $c_9$ & -499.105 \\
    $c_{10}$ & 121.773
    \end{tabular}
    \caption{Numerical values of the coefficients $c_n$ appearing in Eq. \eqref{eq:Ffnlo_polynom}.
    }
    \label{tab:cn_coeff}
\end{table}

The one-loop formula has an explicit dependence on the renormalization scale $\mu$ in the form of  $\log (\mu^2r^2)$ and an implicit dependence through the perturbative strong coupling $\alpha_s(\mu)$. 
At zero flow time, the only scale is the distance $r$; therefore, setting $\mu=1/r$ is a natural choice.
At large flow time, $r$ is negligible with respect to $\sqrt{8\tau_F}$, and therefore, 
the natural choice is $\mu=1/\sqrt{8\tau_F}$. A parametrization of $\mu$ that interpolates between these two limiting cases is $\mu=1/\sqrt{r^2+8\tau_F}$ \cite{Brambilla:2021egm}.
Because in our lattice calculations we are not collecting data at large flow time, 
we adopt in this work the more general parametrization
\begin{align}
    \mu(r,\tau_F) = \frac{1}{\sqrt{sr^2+8b\tau_F}}
    \label{eq:generic_mu_b}\,.
\end{align}
At zero flow time, $\mu(r,0) =  1/(\sqrt{s}r)$, and we can interpret $\sqrt{s}$ as a scale-variation parameter with central value 1. At intermediate-flow-time values i.e., $\sqrt{8\tau_F}$ of the order of $r$
the parameter $b$ defines an effective flowed distance.
Starting from three loops,  an ultrasoft (us) scale of order $\alpha_\mathrm{s}/r$ 
also enters the static force equations at zero flow time~\cite{Brambilla:1999qa}. 
For the scope of this paper, we set the ultrasoft scale to be
$\mu_\mathrm{us}=C_\mathrm{A}\alpha_\mathrm{s}(1/r)/(2r)$.

We renormalize the coupling in the $\overline{\textrm{MS}}$ scheme; hence, both $\alpha_s$ and the scale 
$\Lambda_{\overline{\textrm{MS}}}$ are defined in that scheme.
The scale $\Lambda_{\overline{\textrm{MS}}}$ can be obtained by comparing the perturbative expression of the force with lattice data.
Since we work in the pure SU(3) theory, the comparison provides $\Lambda_0\equiv\Lambda_{\overline{\textrm{MS}}}^{n_f=0}$.

The small-flow-time expansion of $r^2F$ reads
\begin{align}
r^2F(r,\tau_F) \approx r^2F(r,\tau_F=0) + \frac{\alpha_s^2C_F}{4\pi}\left[ -12\beta_0 - 6C_A c_L\right] \frac{\tau_F}{r^2}\,,
\label{eq:NLO_force_flowed}
\end{align}
with $c_L=-22/3$. We remark that $[-12\beta_0 - 6C_A c_L]=8n_f$, which is 0 in this study ($n_f=0$). This means that at small flow time, the static force approaches a constant behavior, and that corrections to it are smaller than $\alpha_s^2 \tau_F/r^2$; $r^2F(r,\tau_F=0)$ is $r^2$ times the one-loop perturbative force at zero flow time.

The static force at zero flow time is known up to N$^3$LL accuracy \cite{Brambilla:1999qa,Pineda:2000gza,Brambilla:2006wp,Brambilla:2009bi,Anzai:2009tm,Smirnov:2009fh}, and the higher loop contributions are crucial for the extraction of the $\Lambda_0$ parameter. To benefit from this knowledge, we model the flowed force with the 1-loop expression at finite flow time, and we demand it to converge to the expression at arbitrary order at zero flow time. In terms of equations, our model function is given by
\begin{align}
    r^2F(r,\tau_F) &= r^2F(r,\tau_F=0) + f^{\mathrm{1\mhyphen loop}}(r,\tau_F) \,,\label{eq:flowed_force_given_order} \\
    f^\mathrm{1\mhyphen loop}(r,\tau_F) &= r^2F^\mathrm{1\mhyphen loop}(r,\tau_F) - r^2F^\mathrm{1\mhyphen loop}(r,\tau_F=0)\,,
\end{align}
where $r^2F(r,\tau_F=0)$ is the static force at a given order at zero flow time, and $r^2F^\mathrm{1\mhyphen loop}$ is the full one-loop expression in Eq. \eqref{eq:flowed_force_full_1_loop_formula}. In this way, we correct for the change of the force due to the flow time up to one-loop order. 
The accuracy of the flow-time correction is consistent with the three-loop accuracy of the zero-flow-time part, as long as the flow-time 
correction subleading to $\alpha_s^2 \tau_F/r^2$ is small compared to $\alpha_s^4$. This appears to be the case in our study, where we invoke the restriction $\tau_F/r^2 \lesssim 0.05$. 

For the rest of the paper, we refer to Eq. \eqref{eq:flowed_force_given_order} when dealing with the flowed static force at higher orders. We label the one-loop-order force [next-to leading order (NLO)] as F1l, the two-loop force [next-to-next-to leading order (N$^2$LO)] as F2l, the two-loop force with leading ultrasoft logarithms resummed [next-to-next-to leading logarithmic order (N$^2$LL)] as F2lLus, the three-loop force [next-to-next-to-next-to leading order (N$^3$LO)] with F3l, and the three-loop force with leading ultrasoft logarithms resummed as F3lLus. 
For the reasons discussed in \cite{Bazavov:2014soa}, we also restrict the present study to the F3lLus force, 
although the force at three-loop order with next-to-leading ultrasoft logarithms resummed [next-to-next-to-next-to leading logarithmic order (N$^3$LL)] would be available.

\subsection{The force on the lattice}
The Wilson loops $W_{r\times T}$ are constructed as the closed, path-ordered product of link variables, consisting of two straight spatial Wilson lines in the spatial plane separated by $T$ in the temporal direction. The spatial Wilson lines have the length $r$ in the direction $\hat{\mathbf{r}}$. The ends of both spatial Wilson lines are connected by two straight temporal Wilson lines. The static force can be obtained as the numerical derivative of Eq. \eqref{eq:static_energy_definition} from the symmetric finite difference
\begin{align}
    F_{\partial V} = \frac{V(r+a) - V(r-a)}{2a}.\label{eq:lattice_static_energy_derivative}
\end{align}
Other methods of defining the derivative of Eq.~\eqref{eq:static_energy_definition} consist, for example, in using  the derivative of interpolating functions; these methods, however, add additional systematic uncertainties.

The main purpose of this work is to obtain the static force directly by computing 
$P\,W_{r\times T}gE_j(\mathbf{r}=r\hat{\mathbf{j}},t^*)$, which consists of inserting a discretized $j$th chromoelectric field component into the path-ordered product at the temporal position $t^*$ in one of the temporal Wilson lines. In general, $t^*$ is arbitrary. Nevertheless, we choose $t^*=T/2$ for even-spaced separations, and an average over $t^*=T/2\pm a/2$ for odd-spaced separations. This reduces the interactions between the chromoelectric field and the corners of the Wilson loop. We use the clover discretization for the field-strength tensor,
\begin{align}
    a^2F_{\mu\nu}&=\frac{-i}{8}(Q_{\mu\nu}-Q_{\nu\mu})\,, \label{eq:clover_leaf_definition}\\
    Q_{\mu\nu} &= U_{\mu,\nu}+U_{\nu,-\mu}+U_{-\mu,-\nu}+U_{-\nu,\mu} = Q_{\nu\mu}^\dagger\,,
\end{align}
where $U_{\mu\nu}$ is a plaquette in the $\mu$-$\nu$ plane. This symmetric definition of the chromoelectric field corresponds to the symmetric center difference according to Eq.~\eqref{eq:lattice_static_energy_derivative} at tree level. Finally, we replace $a^2F_{\mu\nu}$ with $a^2F_{\mu\nu}- \mathbb{1} \, \mathrm{Tr}(a^2F_{\mu\nu})/3$,  which makes the components of the field-strength tensor traceless and corresponds to an $a^2$ improvement~\cite{Bilson-Thompson:2002xlt}. The chromoelectric field components are accessible through the components $a^2E_i=-a^2F_{i,4}$.

The direct determination of the force $F_E$ on the lattice follows from  Eq.~\eqref{eq:static_force_E_insertion_definition} and the discretized version of 
$P\,W_{r\times T}gE_j(\mathbf{r}=r\hat{\mathbf{j}},t^*)$. The finite extent of the chromoelectric field through its discretization introduces additional self energy contributions with a nontrivial lattice spacing dependence.
These self-energy contributions slow down the convergence to the continuum limit; they have been studied in lattice perturbation theory~\cite{Lepage:1992xa}. They are absent in the force obtained through the derivative, $F_{\partial V}$. Since both calculations provide the same physical quantity,  we may  set 
\begin{linenomath}
    \begin{align}
        Z_E F_E(r) = F_{\partial V} \,, \label{eq:lattice_E_field_renormalization}
    \end{align}
\end{linenomath}
where the constant $Z_E$ reabsorbs the additional self-energy contributions at finite lattice spacing. 
If $Z_E=1$, no self-energy contributions are present, and we can assume that the quantity behaves in a trivial way in  the continuum limit. $Z_E$ from the static force was investigated nonperturbatively in a former study  \cite{Brambilla:2021wqs}, and it was found that $Z_E$ has only a weak $r$ dependence. For a different discretization of the chromoelectric field insertion needed for determining transport coefficients, the renormalization constant $Z_E$ was computed up to NLO in lattice perturbation theory \cite{Christensen:2016wdo}. More recent studies \cite{Brambilla:2022xbd, Altenkort:2020fgs} rely on the gradient flow method to renormalize the field insertions. In this study, we also use gradient flow to show the renormalization property explicitly, and to improve the signal-to-noise ratio. In the rest of this work, the default force measurement is given in terms of the chromoelectric field -- i.e., $F\equiv F_{E}$ -- while we still call the force obtained from the derivative of the static energy $\partial_r V\equiv F_{\partial V}$.

\section{Lattice setup and technical details}\label{sec:lattice_setup}
On the lattice, we measure $a^2F(t,r,T)$, which, multiplied by $r^2/a^2$, yields the dimensionless quantity $r^2F(t,r,T)$. We have data on a periodic $N_s^3\times N_t$ ($N_s$: spatial lattice extent, $N_t$: temporal lattice extent) grid; in our case, the available grids are $20^3\times 40$, $26^3\times 52$, $30^3\times 60$, and $40^3\times 80$. Table~\ref{tab:lattice_parameters} shows our lattice parameters. We use the scaling from~\cite{Necco:2001xg}, which is based on the scale $r_0$, to fine-tune the simulation parameters. With this scaling, our lattices share a physical size of approximately $(\SI{1.2}{fm})^3\times \SI{2.4}{fm}$.
We produce the lattice configurations using over-relaxation and heat-bath algorithms with Wilson action and periodic boundary conditions.

\begin{table}
    \centering
    \begin{tabular}{c|c|c|c|c|c|c}
        $N_S$ & $N_T$ & $\beta$ & $a$ [fm] & $t_0/a^2$ &  $N_\mathrm{conf}$ & Label \\\hline
        $20^3$ & $40$ & 6.284 & 0.060 & 7.868(8) & 6000 &L20\\\hline
        $26^3$ & $52$ & 6.481 & 0.046 & 13.62(3) & 6000 & L26\\\hline
        $30^3$ & $60$ & 6.594 & 0.040 & 18.10(5) & 6000 & L30\\\hline
        $40^3$ & $80$ & 6.816 & 0.030 & 32.45(7) & 3300 & L40\\\hline
    \end{tabular}
    \caption{The parameters for the lattice ensembles. The scale $t_0$ was set on a smaller subset of lattice configurations.}
    \label{tab:lattice_parameters}
\end{table}

We solve the gradient flow equation~\eqref{eq:gradient_flow_definition_lattice} with a fixed step-size integrator for the coarsest lattice ($20^3\times 40$) and an adaptive step-size integrator for the finer lattices. In all gradient flow integrations, we use the Symanzik action. We measure pure Wilson loops and loops with chromoelectric field insertions. That way, we can determine both the static potential and its numerical derivative for obtaining the static force, and the force directly. In Appendix \ref{sec:flowed_wilson_loops}, we show briefly the impact of gradient flow on the bare Wilson loops with and without chromoelectric field insertions. To find the reference scale $t_0$, we use the clover discretization in Eq.~\eqref{eq:clover_leaf_definition} to measure $\langle E\rangle$ and solve Eq. \eqref{eq:t0_definition} for $t_0$. We use this reference scale to express our quantities $r$ and $\tau_F$ in units of $\sqrt{t_0}$ and $t_0$, respectively, and to perform the continuum limit as $a^2/t_0\rightarrow 0$.

The adaptive step-size integrator changes the gradient flow step sizes after every integration step, dependent on the lattice configuration. This means that the lattice measurement is done at different flow-time grids for each individual lattice configuration. Therefore, we need to interpolate the data to a common flow-time grid among the different lattice configurations. We use simple spline interpolations, since the gradient flow is a continuous transformation of the fields that produces a continuous function of the flow time.
The interpolation can be done to a fixed flow-time grid in physical units, or to a fixed flow-time-ratio $\tau_F/r^2$ grid. 
Data along a flow-time grid in physical units at a given fixed $r$ can be easily presented on a flow-time-ratio axis by setting the $x$ axis to $\tau_F/r^2$.

We check the fluctuation of the topological charge and observe no full freezing of the topology. At our largest lattice (L40), the fluctuation of topology slows down, which increases the autocorrelation times of the topological charge. We inspect the autocorrelation of the L40 lattice more closely in Appendix~\ref{app:autocorrelation}. The static energy (and consequently, its derivative) is known to be less affected by topological slowing down~\cite{Weber:2018bam} than a bare gradient-flow-coupling measurement would be~\cite{Fritzsch:2013yxa}. However, to be safe, we block our data such that the block size is larger than the topological autocorrelation times we see on any of the ensembles and settle for 30 jackknife blocks per ensemble.
While the static force measurement is only weakly correlated with the topological charge, the scale $t_0$ can have a stronger dependence on it. To counter this, we measure $t_0$ only on a subset of configurations, having longer Monte Carlo time in between the measurements.

We have performed the simulations at a constant physical box size and have not tested for finite-volume effects from varying the physical volume. 
In~\cite{Schlosser:2021wnr}, the finite-volume effects were studied and found to be minimal 
for hybrid static energies for almost the same set of lattice parameters. 
We expect the finite-volume effect to be equally small for the static force. 

\section{Preparatory analyses}\label{sec:prep_analysis}
In this section, we analyze and prepare the raw lattice data, and we perform the continuum limits needed for the $\Lambda_0$ extraction. This preparation contains the plateau extraction for the static energy and the $T\rightarrow\infty$ limit for the direct force measurement, which is covered in Sec. \ref{sec:plateau_extraction}. Renormalization properties of gradient flow are discussed in Sec. \ref{sec:implications_of_gradientflow}, followed by the continuum extraction, worked out in Sec.~\ref{sec:continuum_extrapolation}. To finalize this section, we investigate the behavior of the various scales $r_0$, $r_1$, and $t_0$.

\subsection{Plateau extraction}\label{sec:plateau_extraction}
We extract the plateaus in the $T\rightarrow \infty$ limit based on a procedure from~\cite{Jay:2020jkz} that relies on an Akaike information criterion. 
We summarize it here briefly. This procedure is applied to data at every fixed $r$ and $\tau_F$ combination for each lattice ensemble.

In the first step, we perform constant fits for all possible continuous ranges within $T/a=1$ and a certain $T_\mathrm{max}$ with a minimum of at least three support points, each fit minimizing $\chi^2$ as
\begin{align}
    \mathbf{a}^\star_{i_1,i_2} = &\ \underset{\mathbf{a}_{i_1,i_2}}{\mathrm{argmin}}\ \chi^2(\mathbf{a}_{i_1,i_2}) = \underset{\mathbf{a}_{i_1,i_2}}{\mathrm{argmin}} \sum _{i,j=i_1}^{i_2} \left( f(x_i,\mathbf{a}_{i_1,i_2}) - y_i \right) C_{i,j}^{-1}\left( f(x_j,\mathbf{a}_{i_1,i_2}) - y_j \right),
\end{align}
where $i_1$ and $i_2$ are the indices defining the specific lower and upper limits of the range; $f$ represents the model function, which, in our case, is the constant function $f(x,\mathbf{a})=a_0$; and $\mathbf{a}$ is the parameter vector, which consists of only one component for the constant fit. The dataset $D$ is specified by $y_i$, the force measurement at $T=x_i$, and $C_{ij}$ is the covariance matrix along the $T$ axis.

In the next step, we define, for a given fit and for a specific range $i_1$ to $i_2$, the Akaike information criterion (AIC) as
\begin{align}
    \mathrm{AIC}_{i_1,i_2} = \chi^2(\mathbf{a}^\star_{i_1,i_2}) + 2k + 2(i_1 - i_2)\,,\label{eq:AIC_definition}
\end{align}
where $k$ is the number of parameters of the fit function, $k=1$ for the constant fit, and $i_1-i_2$ is the relative number of discarded data points within the total range. In theory, we are required to have the absolute number of discarded data points, which is given by $N_\mathrm{tot} - (i_1-i_2)$; however, this corresponds to a global shift for all $\mathrm{AIC}$ values, which can be eliminated in the next step.

In the third step, we find the model probability of a specific fit range as
\begin{align}
    p(i_1, i_2 | D) = \frac{1}{Z} \exp^{-\frac{1}{2}\mathrm{AIC}_{i_1,i_2}}\,,\label{eq:AIC_probability}
\end{align}
where $Z$ is a normalization constant such that the sum of the probabilities over all considered fit ranges reduces to 1:
\begin{align}
    \sum _{i_1<i_2} p(i_1,i_2|D) = \frac{1}{Z} \sum _{i_1<i_2} \exp^{-\frac{1}{2}\mathrm{AIC}_{i_1,i_2}} \overset{!}{=} 1.
\end{align}
At this stage, we observe that a global shift of all AIC values has no effect on the model probability, since it is absorbed into a redefinition of $Z$. Thus, we perform the global shift  $\mathrm{AIC}_{i_1,i_2}\rightarrow \mathrm{AIC}_{i_1,i_2} - \underset{i_1,i_2}{\mathrm{min}}\ \mathrm{AIC}_{i_1,i_2}$, because it makes the computations of the model probability on the most important ranges numerically more stable.

In the last step, we compute model expectation values and deviations with the given model probabilities. The final results and their deviations from the plateau fit are thus given by
\begin{align}
    \Bar{a}_n &= \langle a_n \rangle = \sum_{i_1,i_2} a^\star_{n,i_1,i_2}p(i_1,i_2|D) \,,\\
    \sigma _n^2 &= \langle (a_n - \Bar{a}_n)^2\rangle = \langle a_n^2\rangle - \langle a_n \rangle ^2.\label{eq:aic_deviation_definition}
\end{align}
This procedure can be generalized to more complex model functions, which we use in this work for the continuum limit and the $\Lambda_0$ extraction. Furthermore, to achieve a better understanding of the process, we compute the average fit ranges
\begin{align}
    \langle x_{i_1} \rangle &\equiv \sum_{i_1,i_2} x_{i_1} p(i_1,i_2|D) ,\label{eq:AIC_lower_fit_range}\\
    \langle x_{i_2} \rangle &\equiv \sum_{i_1,i_2} x_{i_2} p(i_1,i_2|D), \label{eq:AIC_upper_fit_range}
\end{align}
and their deviations, defined equivalently to Eq.~\eqref{eq:aic_deviation_definition}.

A crucial part of this procedure is the selection of a certain $T_\mathrm{max}$. In principle, $T_\mathrm{max}$ can be chosen to cover the whole temporal range of the Wilson loops, since the information criterion should eventually select the important ranges. However, especially at larger $T$, the statistical errors for the covariance matrix are underestimated. These give inaccurate model probabilities for insignificant fit ranges. Therefore, we have to select a proper $T_\mathrm{max}$ value to prevent this behavior. To find a suitable $T_\mathrm{max}$, we perform the whole procedure for several $T_\mathrm{max}$ values and select the one where, for a small variation of $T_\mathrm{max}$, the final result stays invariant. This is justified by the assumption that the important fit ranges, selected by the information criterion, should be fully captured by $T_\mathrm{max}$. 
Hence, a small variation of $T_\mathrm{max}$ should not modify the important fit ranges. The $T_\mathrm{max}$ selection can be automatized by taking data up to a $T_\mathrm{max}$ where the relative statistical error is less than \SI{10}{\percent}, and then decreasing $T_\mathrm{max}$ iteratively until an invariant range of $T_\mathrm{max}$ is found. However, the automatized procedure does not work in every case, and sometimes a selection has to be made manually. To do so, it is enough to identify a representative selection of a few flow times and distances $r$, and use piecewise linear interpolations to cover the whole dataset.

The Akaike information procedure also provides us an error estimate of the plateau extraction. Nevertheless, to propagate the statistical error, we use jackknife resampling with 30 blocks and perform the plateau extraction for every jackknife block. 
If the resulting jackknife error is comparable with the fit error, we use only the jackknife samples to propagate the error. In several cases, however, the choice of the fit range is a significant systematic error source, and we need to take it into account. In these cases, we label the systematic error by the Akaike information criterion of the corresponding observable as $\sigma^2_\mathrm{AIC}$, and the statistical error as $\sigma^2_\mathrm{stat}$.

\subsection{Implications of the gradient flow}\label{sec:implications_of_gradientflow}
\begin{figure}
    \centering
    \includegraphics[width=0.6\textwidth]{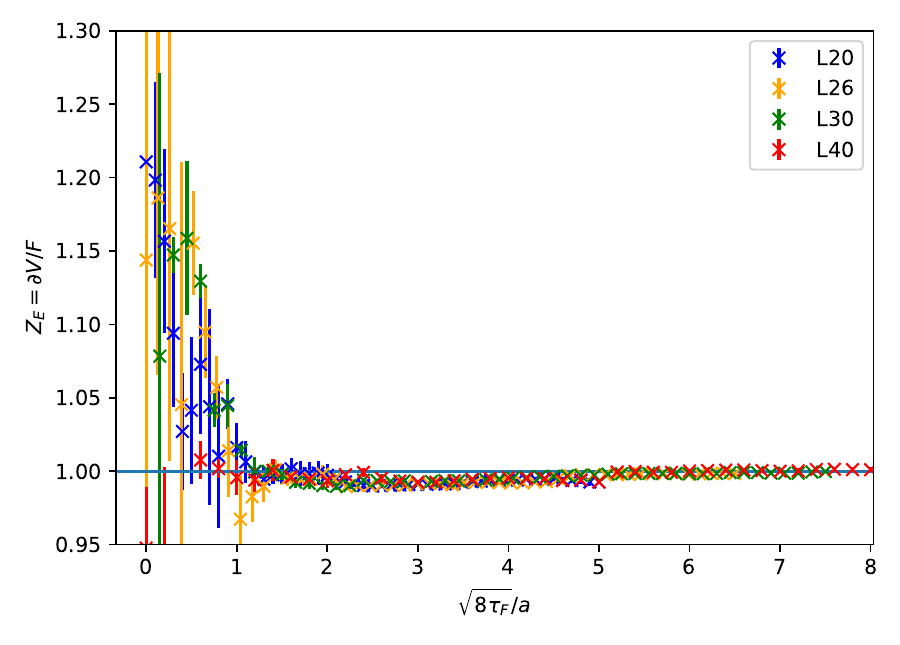}
    \caption{$Z_E$ for all lattice sizes with increasing flow time. We see that for flow radii larger than one lattice spacing, the factor $Z_E$ becomes practically 1.}
    \label{fig:Ze_joint}
\end{figure}

Gradient flow has an impact not only on the signal-to-noise ratio improvement, but also on reducing discretization effects that occur through self-interaction contributions. Therefore, we are also interested in the renormalization factor $Z_E$ of the chromoelectric field insertion from Eq.~\eqref{eq:lattice_E_field_renormalization}. We find $Z_E$ nonperturbatively by solving Eq.~\eqref{eq:lattice_E_field_renormalization} for $Z_E$, which gives the ratio of the numerical derivative of the static potential and the direct force measurement according to Eq. \eqref{eq:static_force_E_insertion_definition}:
\begin{align}
    Z_E(r)=\frac{\partial _rV(r)}{F(r)}.
\end{align}
In \cite{Brambilla:2021wqs}, it was shown that $Z_E(r)$ has a weak $r$ dependence. 
Hence, we  extract $Z_E$ as a plateau fit, similar to the $T\rightarrow\infty$ limit discussed in Sec.~\ref{sec:plateau_extraction}, over $r$ at fixed $\tau_F$; we keep $r$ approximately between $0.3\,r_0$ and $0.65\,r_0$.
Figure~\ref{fig:Ze_joint} shows the result for $Z_E$ for all lattice sizes against the flow time. At minimal flow times, we obtain that $Z_E>1$, meaning that at minimal flow times, the direct force measurement is affected by self-energy contributions originating from the chromoelectric field discretization. Furthermore, we obtain that $Z_E= 1$ within \SI{1}{\percent} deviation for flow radii larger than one lattice spacing, which is required for reliable continuum limits. 
We recognize a small bump within the \SI{1}{\percent} range for flow radii within $1.5<\sqrt{8\tau_F}/a<5$ for all lattice sizes. This is expected, as the $\partial _rV(r)$ part is a simple finite difference and hence only approximates the static force. We observe that this systematic difference between the definitions of the force seems to vanish at larger flow radii ($\sqrt{8\tau_F}/a>6$).
In conclusion, we find that a minimum amount of flow time ($\sqrt{8\tau_F}>a$) has to be applied to be in the regime where the gradient flow has practically eliminated the nontrivial discretization effects.

\subsection{Continuum extrapolation}\label{sec:continuum_extrapolation}
\subsubsection{Continuum interpolation}
Preparing for the continuum limit, we interpolate $r^2F$ on all ensembles at a fixed flow time to a common $r$ range in $\sqrt{t_0}$ units.
We use polynomial interpolations at different orders:
\begin{equation}
    P_n (r) = \sum _{k=0}^n a_{k,n} r^k\,,
\end{equation}
with $a_{k,n}$ being the coefficients. We have different coefficients for different fixed-order polynomials. An interpolation at fixed order corresponds to a single fit, where we minimize the $\chi^2$.

In addition, interpolations have the advantage that small fluctuations within the data get smoothed. Polynomial interpolations can have fluctuations, especially higher-order polynomial fits, and hence, we average over different polynomial orders to reduce those fluctuations:
\begin{align}
    P_\mathrm{inter}(r) &= \sum _{n=n_\mathrm{min}}^{n_\mathrm{max}} w_n P_n(r) \,,\\
    \sum _{n=n_\mathrm{min}}^{n_\mathrm{max}} w_n &= 1\,,
\end{align}
where $w_n$ are normalized, adjustable weights.

For the L20, L26, and L30 lattices, we use equal weights for all polynomial orders. We choose polynomials of orders $n=4,5,6,7$ for L20; orders $n=7,8,9,10$ for L26; and orders $n=8,9,10,11$ for L30. For L40, we use an Akaike average \cite{Jay:2020jkz} over the orders from 4 to 12, because at some flow times the plateau extraction at larger $r$ gives fluctuating results. This results from underestimated systematic effects and causes a change in the effective polynomial orders, which is considered through the Akaike average. The weights are found analogously to Eqs.~\eqref{eq:AIC_definition} and \eqref{eq:AIC_probability} through
\begin{align}
    \mathrm{AIC}_n &= \chi^2_n + 2 \,\mathrm{d.o.f.}\,, \label{eq:AIC_general_definition}\\
    w_n &\propto e^{-\frac{1}{2}\mathrm{AIC}_n}\,, \label{eq:AIC_general_weight}
\end{align}
where the final values of $w_n$ are fixed by the normalization condition. To propagate the statistical error, the continuum interpolation is done for every jackknife block.

This procedure works for most of the data. An exception is the data at small $r$ (up to $r/a=3$) for the L26 lattice. We obtain a miscarried interpolation in this case due to large-$r$ effects in the interpolation. 
For the L26 lattice, it turns out that spline interpolations up to $r/a=3$ and changing to the polynomial interpolation for larger $r$ works properly.

\subsubsection{Tree-level improvement}
To reduce the effects of finite lattice spacings, we apply a tree-level improvement procedure to the data at finite flow time by dividing out the leading lattice perturbation theory result. 
Following~\cite{Fodor:2014cpa}, the static energy in lattice perturbation theory at finite flow time can be written as
\begin{equation}
    V_\mathrm{lat}(r,\tau_F) = -C_F g^2 \int \frac{\mathrm{d}^3 \mathbf{k}}{(2\pi)^3} e^{i\mathbf{k}\cdot\mathbf{r}} e^{-2\tau_F D^{-1}_\mathrm{GF}}D_\mathrm{MC}\,,
\end{equation}
where we have assumed $a=1$ for simplicity and $D$ is the lattice propagator:
\begin{equation}
    D^{-1} = 4\sum_{i=1}^4 \left(\sin^2\frac{k_i}{2} + c_w\sin^4\frac{k_i}{2}\right)\,,
\end{equation}
with $c_w=0$ for the Wilson action, which we use for the simulation part, $D_\mathrm{MC}$, 
and $c_w=1/3$ for the Symanzik action, which we use in the gradient flow equations, $D_\mathrm{GF}$.
Similarly to the zero-flow-time case that was derived in~\cite{Brambilla:2021wqs}, the static force coming from the chromoelectric field insertion
reduces to a symmetric finite difference:
\begin{equation}
    F_\mathrm{lat}(r,\tau_F) = \frac{V_\mathrm{lat}(r+a,\tau_F)-V_\mathrm{lat}(r-a,\tau_F)}{2a}\,.  
\end{equation}
Now, we can tree-level-improve the measured force $F_\mathrm{meas}$ by dividing out the leading lattice expression and multiplying by the continuum tree-level gradient flow expression  
\begin{align}
    F_\mathrm{improved}(r,\tau_F) = \frac{F_\mathrm{cont}^\mathrm{tree\mhyphen level}(r,\tau_F)}{F_\mathrm{lat}(r,\tau_F)} F_\mathrm{meas}(r,\tau_F)\,,
\end{align}
where $F_\mathrm{cont}^\mathrm{tree\mhyphen level}$ is defined as the $\mathcal{O}(\alpha_\mathrm{S})$ contribution to Eq.~\eqref{eq:flowed_force_full_1_loop_formula}.

\subsubsection{Continuum extrapolation}

\begin{figure}
\centering
    \includegraphics[width=0.47\textwidth]{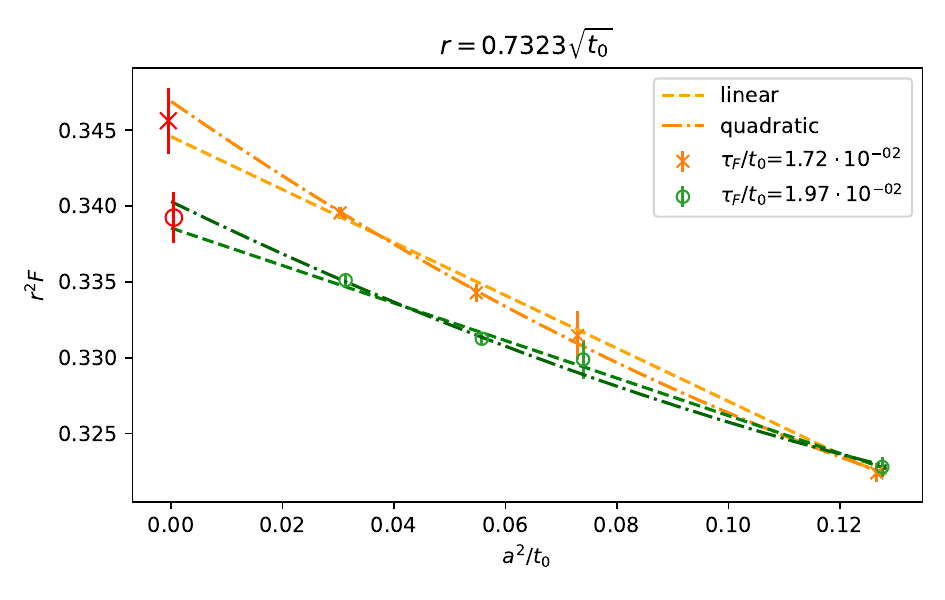}
    \includegraphics[width=0.47\textwidth]{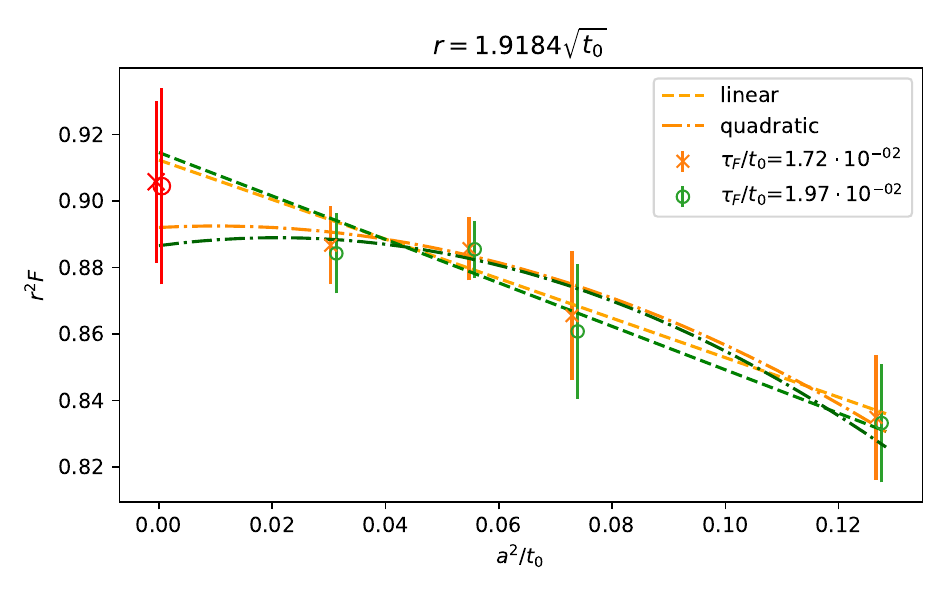}
\caption{Examples of continuum extrapolation for two different separations $r$. The figure shows the linear and quadratic continuum limits: the line corresponds to the linear limit, and the curve to the quadratic one.}
\label{fig:continuum_limit_example}
\end{figure}

We use the interpolated and tree-level improved data for the continuum limit, which is obtained from  extrapolations linear and quadratic in $a^2/t_0$ at fixed physical distances $r$ and fixed physical flow times $\tau_F$:
\begin{align}
    f^\mathrm{linear}(a^2/t_0) &= m \frac{a^2}{t_0} + c_l\,,\\
    f^\mathrm{quadratic}(a^2) &= A \frac{a^4}{t_0^2} + B \frac{a^2}{t_0} + C_l\,,
\end{align}
where $m, c_l, A, B$, and $C_l$ are the fit parameters, and $c_l$ and $C_l$ are the continuum limits of the linear and quadratic extrapolations, respectively. We take an Akaike average, defined for polynomials as in Eqs. \eqref{eq:AIC_general_definition} and \eqref{eq:AIC_general_weight}. Figure \ref{fig:continuum_limit_example} shows a working example for the continuum limit at two fixed distances $r$ and for each distance at two different flow times.

Although we use tree-level improved data and Akaike average over linear and quadratic continuum limits, often both $\chi^2/\mathrm{d.o.f.}$'s are too large. Hence, we restrict to data which accomplish that at least one of the extrapolations gives $\chi^2/\mathrm{d.o.f.}<3.0$ and that the flow radius fulfills $\sqrt{8\tau_F}>a$. The remaining, filtered data represent reliable continuum limit results, with which we continue the further analyses.

\subsection{$r_0$ and $r_1$ scales}\label{sec:r0_r1_scales}

\begin{figure}
\centering
\begin{subfigure}{.45\textwidth}
    \centering
    \includegraphics[width=\textwidth]{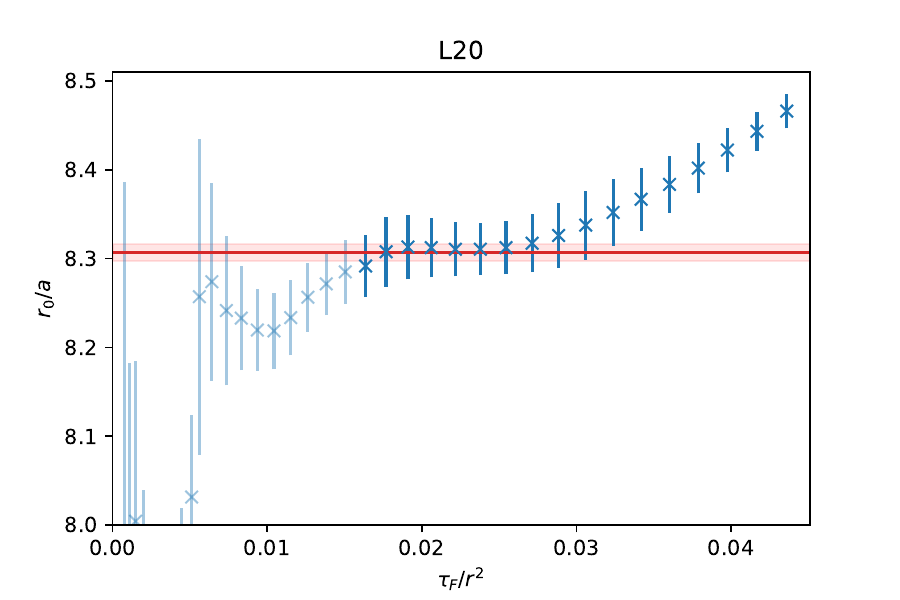}
\end{subfigure}
\begin{subfigure}{.45\textwidth}
    \centering
    \includegraphics[width=\textwidth]{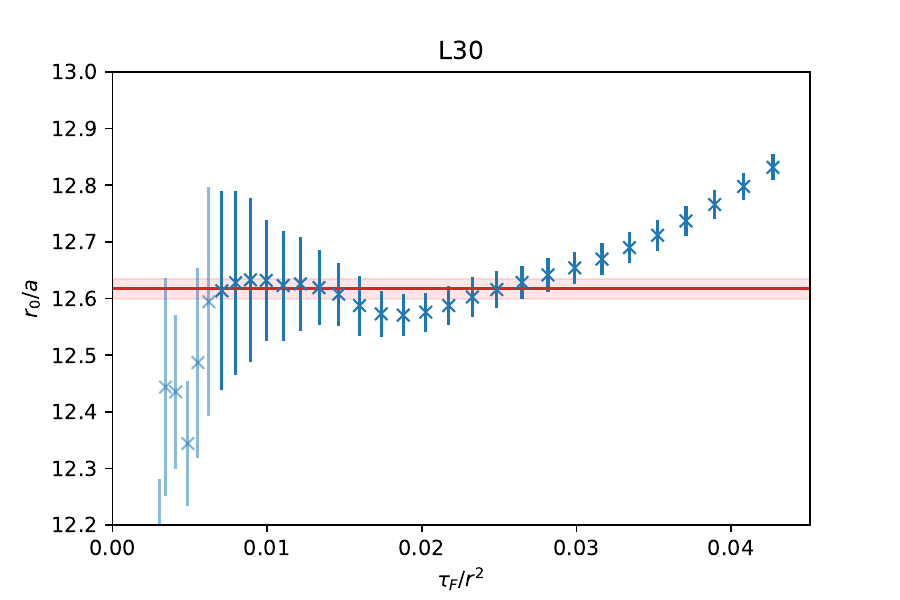}
\end{subfigure}
\caption{The $r_0$ scale along the flow-time axis.}
\label{fig:r0_extraction}
\end{figure}

\begin{figure}
\centering
\begin{subfigure}{.45\textwidth}
    \centering
    \includegraphics[width=\textwidth]{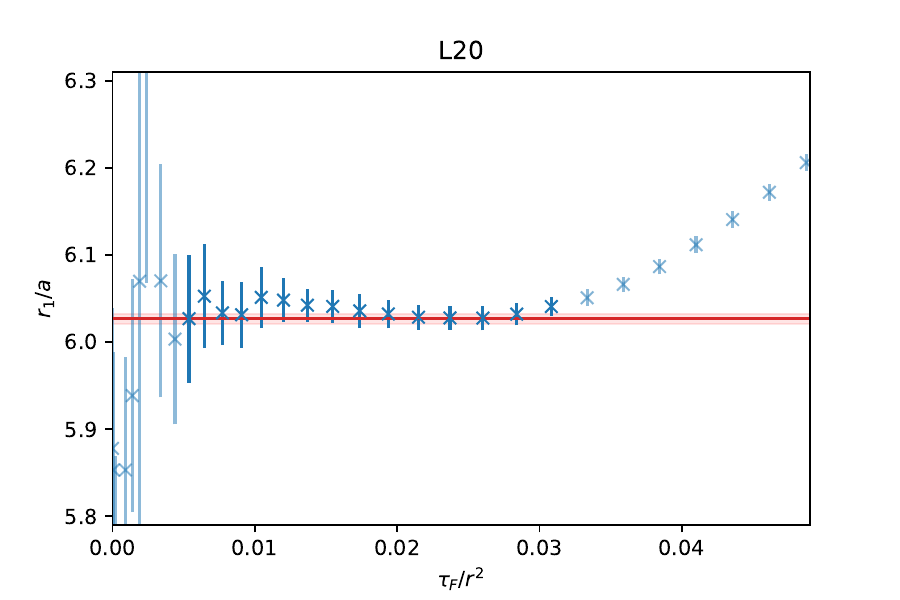}
\end{subfigure}
\begin{subfigure}{.45\textwidth}
    \centering
    \includegraphics[width=\textwidth]{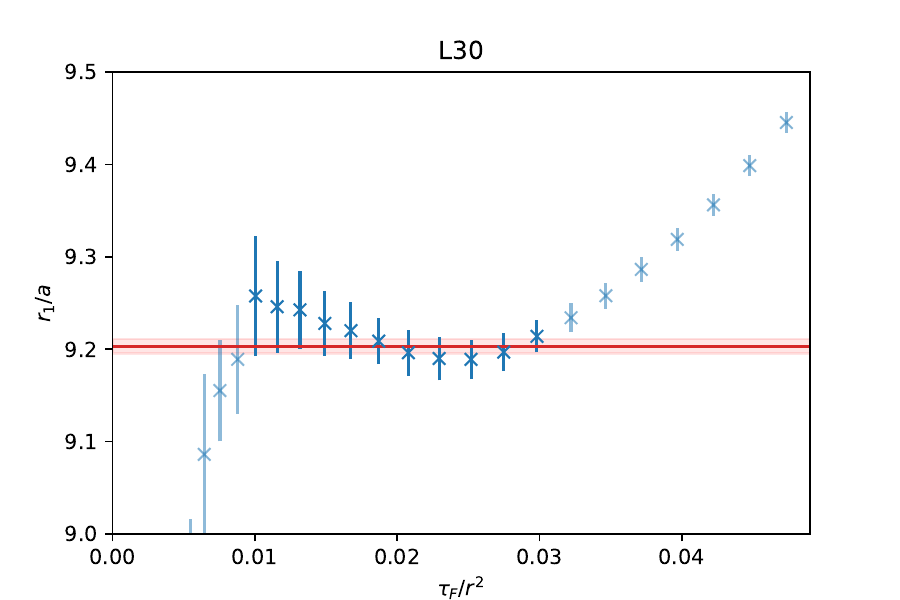}
\end{subfigure}
\caption{The $r_1$ scale along the flow-time axis.}
\label{fig:r1_extraction}
\end{figure}

\begin{table}[]
    \centering
    \begin{tabular}{l|c|c|c|c||c|c|c|c}
                    & $r_0/a$ & $\sigma_\mathrm{stat}$ & $\sigma_\mathrm{AIC}$ & $\sigma_\mathrm{stat+AIC}$ & $r_1/a$ & $\sigma_\mathrm{stat}$ & $\sigma_\mathrm{AIC}$ & $\sigma_\mathrm{stat+AIC}$\\\hline
         L20          & 8.306    & 0.017   & 0.010   & 0.020 & 6.026    & 0.0090   & 0.006   & 0.011\\
         L26          & 10.833   & 0.031   & 0.025   & 0.040 & 7.849    & 0.014   & 0.010   & 0.018\\
         L30          & 12.617   & 0.032   & 0.018   & 0.040 & 9.203    & 0.0130   & 0.008   & 0.015\\
         L40          & 16.933   & 0.042  & 0.015   & 0.044 & 12.316   & 0.0280   & 0.011   & 0.030
    \end{tabular}
    \caption{Results for $r_0$ and $r_1$ 
   in the  zero-flow-time limit at finite lattice spacing. The table includes the statistical error (``stat" subscript), the systematic error by choosing different fit ranges (``AIC" subscript), and the errors added in quadrature.}
    \label{tab:zftl_r0_r1_finite_lattice}
\end{table}

\begin{figure}
    \centering
    \includegraphics[width=0.45\textwidth]{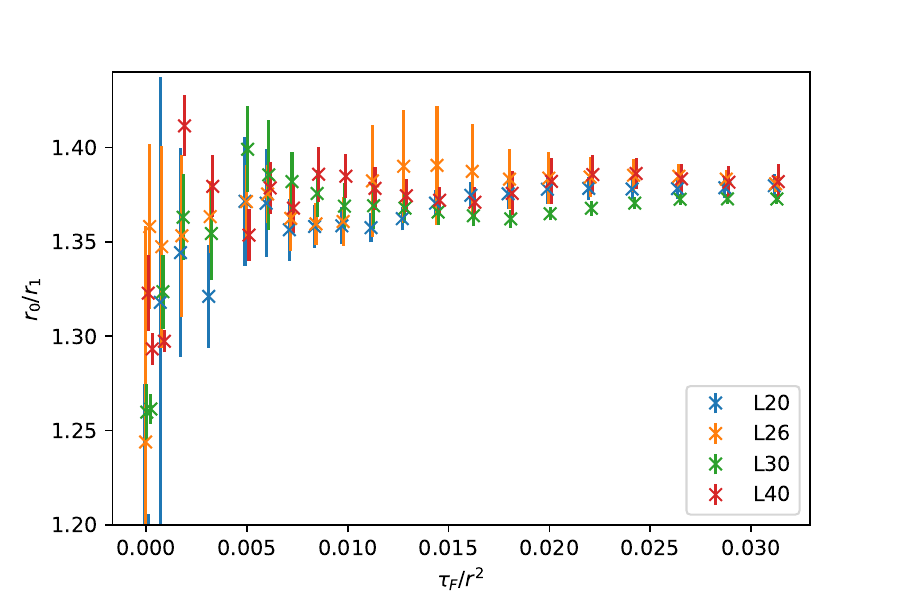}
    \includegraphics[width=0.45\textwidth]{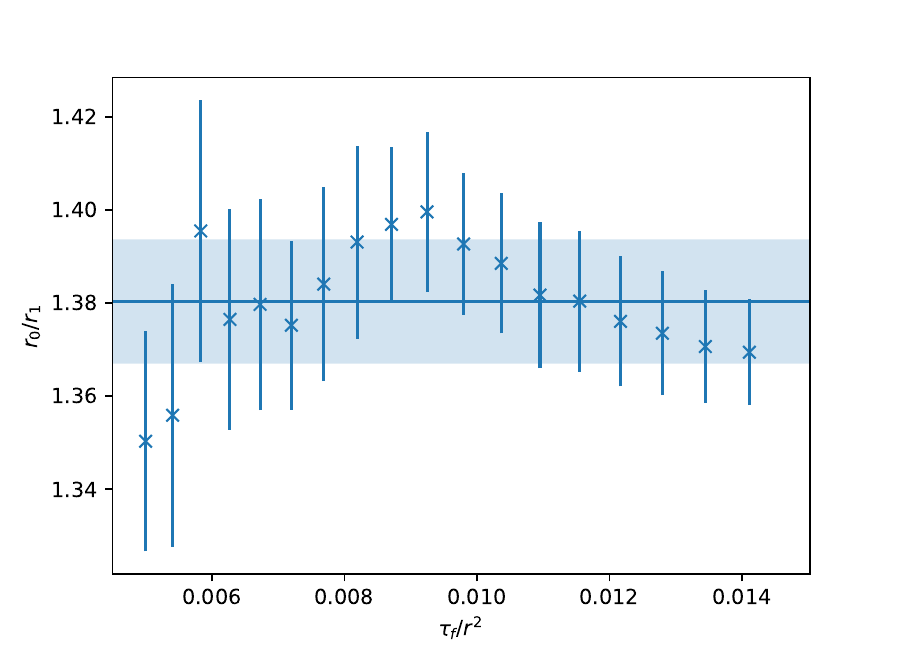}
    \caption{The ratio of the scales $r_0$ and $r_1$. The left side shows the ratio at finite lattice spacing; the right side shows the continuum limit with the constant zero-flow-time extraction.}
    \label{fig:r0_r1_ratio}
\end{figure}

\begin{figure}
    \centering
    \includegraphics[width=0.45\textwidth]{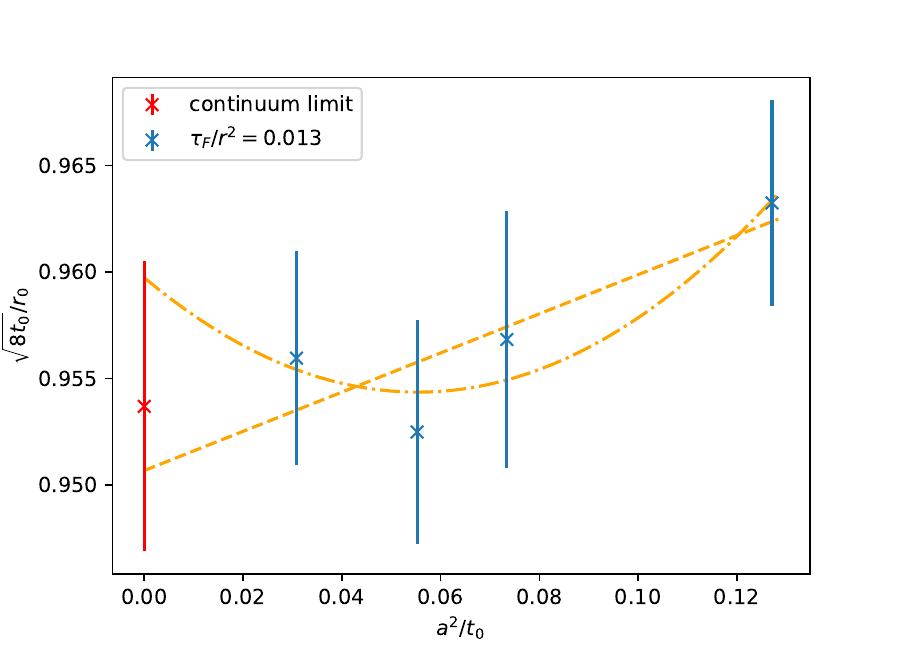}
    \includegraphics[width=0.45\textwidth]{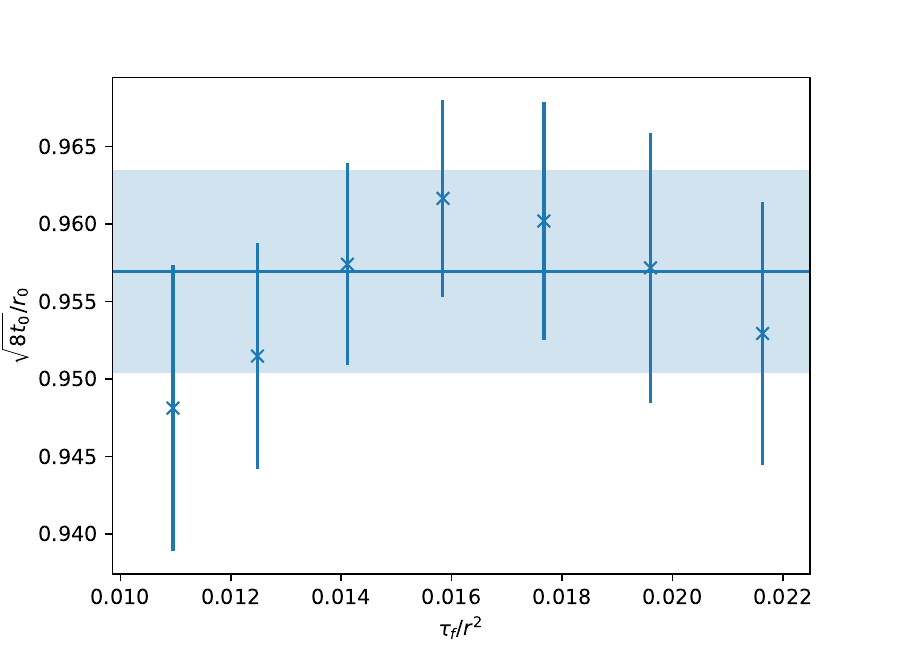}
    \caption{Left: An example for the continuum limit of the ratio $\sqrt{8t_0}/r_0$ at a fixed flow-time ratio. The straight line corresponds to a continuum limit linear in $a^2/t_0$, and the curved line corresponds to a polynomial up to quadratic order in $a^2/t_0$. The red cross and bar shows the final continuum limit of a weighted average of both extrapolations. Right: The flowed continuum ratio with the constant zero-flow-time limit.}
    \label{fig:sqrt8t0r0_ratio_continuum_limits}
\end{figure}

In terms of the force $F(r)$, we define a reference scale as
\begin{equation}
    r^2F(r,\tau_F)|_{r(c,\tau_F)} = c\,,
\end{equation}
with a dimensionless number $c$.
Common choices are $r_0(\tau_F)\equiv r(c=1.65,\tau_F)$, $r_0\equiv r_0(\tau_F=0)$~\cite{Sommer:1993ce}, and $r_1(\tau_F)\equiv r(c=1,\tau_F)$, $r_1\equiv r_1(\tau_F=0)$~\cite{Bernard:2000gd}. In the original proposals, the scales $r(c)$ are defined with the force at zero flow time, however, in our case, the force is computed at different flow times. Thus, we obtain flow-time-dependent $r_0$ and $r_1$ as shown in Figs.~\ref{fig:r0_extraction} and \ref{fig:r1_extraction}. To find $r_0$ and $r_1$, we perform multiple polynomial interpolations of $r^2F(r,\tau_F)$ for larger $r$ values along the $r$ axis and at fixed flow-time ratio $\tau_F/r^2$, and we find the roots of the individual interpolations as $(r^2F)^\mathrm{inter } - c = 0$. The final scales are given in terms of an Akaike average over the roots of the polynomial interpolators.

Both scales, $r_0$ and $r_1$, approach a constant plateau within a recognizable flow-time regime and start deviating from this plateau at larger flow times. We perform plateau fits within this range for a zero-flow-time extraction of both scales, and we find them comparable to the scales fixed by~\cite{Necco:2001xg}. The results are shown in Table~\ref{tab:zftl_r0_r1_finite_lattice}. We see that the error of the scale setting is dominated by the statistical fluctuation rather than by the Akaike error of the polynomial interpolators.

The ratio of the scales, $r_0(\tau_F)/r_1(\tau_F)$, is shown in Fig.~\ref{fig:r0_r1_ratio}. We take the $a^2/t_0$ continuum limit followed by a constant zero-flow-time limit. Finally, we obtain
\begin{equation}
    \frac{r_0}{r_1} = \num{1.380(14)}.
\end{equation}
To our knowledge, there are no prior direct determinations of the scale ratio $r_0/r_1$ in pure gauge. 
However, in Ref.~\cite{Sommer:2014mea} this ratio was roughly estimated 
to be $\sim 1.37$ based on the curves shown in~\cite{Necco:2001xg}, which agrees well with our extraction within errors. 
The extracted value is about 9\% larger than
the ratio in unquenched theories with 2+1 or 2+1+1 fermion flavors~\cite{FlavourLatticeAveragingGroupFLAG:2021npn,Brambilla:2022xbd}. 
Such a shift is to be expected, due to the effect of unquenching the quark flavors. A similar effect between quenched and unquenched scales has been seen for the gradient-flow scale ratios in Ref.~\cite{Bruno:2013gha}.

\begin{figure}
    \centering
    \includegraphics[width=0.45\textwidth]{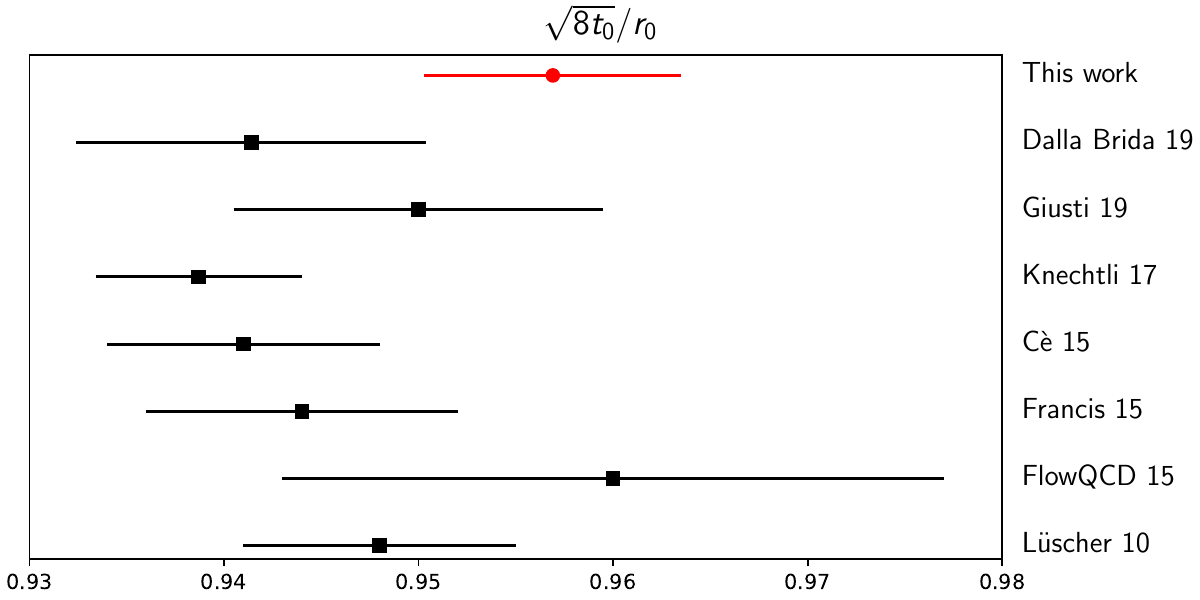}
    \caption{Our extracted ratio (red circle) of the scale ratio $\sqrt{8t_0}/r_0$ compared to the existing results (black squares) in the literature.}
    \label{fig:t0r0_ratio_lit}
\end{figure}

We repeat the same procedure for the ratios $\sqrt{8t_0}/r_0$ and $\sqrt{8t_0}/r_1$. Figure \ref{fig:sqrt8t0r0_ratio_continuum_limits} shows an example for the ratio $\sqrt{8t_0}/r_0$. The final results for the ratios of the scales are
\begin{align}
    \frac{\sqrt{8t_0}}{r_0} &= \num{0.9569(66)} \,,\label{eq:sqrt8t0r0_result}\\
    \frac{\sqrt{8t_0}}{r_1} &= \num{1.325(13)}\,.\label{eq:sqrt8t0r1_result}
\end{align}
Our extracted ratio $\sqrt{8t_0}/r_0$ agrees within errors with the previous determinations~\cite{Luscher:2010iy,Asakawa:2015vta,Francis:2015lha,Ce:2015qha,Kamata:2016any,Knechtli:2017xgy,Giusti:2018cmp,DallaBrida:2019wur},\footnote{The FlowQCD result from Ref.~\cite{Asakawa:2015vta} is inferred from their results of $\sqrt{t_0}/w_0$ and $r_0/w_0$ neglecting error correlation. Therefore, the error shown in  Fig.~\ref{fig:t0r0_ratio_lit} is certainly overestimated.}
albeit the mean value is slightly above most of the existing results.
We show the comparison to the existing literature in Fig.~\ref{fig:t0r0_ratio_lit}. 
The previous results are somewhat correlated, since most of them focus on $t_0$ calculation and use the data from Refs.~\cite{Guagnelli:1998ud,Necco:2001xg} for at least part of their dataset for $r_0$. 
Again, the quenched ratios are larger than the unquenched ones~\cite{FlavourLatticeAveragingGroupFLAG:2021npn} as 
has been previously seen in Ref.~\cite{Bruno:2013gha}. To our knowledge, this is the first direct measurement of the scale ratio $\sqrt{8t_0}/r_1$ in pure gauge.

\section{Analyses of the continuum results}\label{sec:continuum_analysis}
After having worked out the continuum limit in Sec.~\ref{sec:continuum_extrapolation}, we compare the lattice results with the perturbative expressions to extract  $\Lambda_0$. Since gradient flow introduces another scale next to $1/r$, we have several possibilities to compare the lattice results with the perturbative expressions. In the first approach, we use the simple expression of the flowed force in Eq. \eqref{eq:NLO_force_flowed}, which turns out to be also applicable to large-$r$ results. In the second approach, we compare the lattice results either by keeping the scale $1/r$ fixed and inspecting the behavior along the flow time, or vice versa, by keeping $\tau_F$ fixed and inspecting the behavior along the distance $r$. 
We show plots only for the perturbative one-loop (F1l) and three-loop with leading ultrasoft resummation (F3lLus) expressions; however, in the result tables, we also provide extractions based on the two-loop with and without leading ultrasoft resummation (F2l and F2lLus), and based on three-loop without ultrasoft resummation (F3l) expressions.

\subsection{Constant zero-flow-time limit at fixed $r$}\label{sec:large_r_constant_zftl}

\begin{figure}
\centering
\begin{subfigure}{.45\textwidth}
    \centering
    \includegraphics[width=\textwidth]{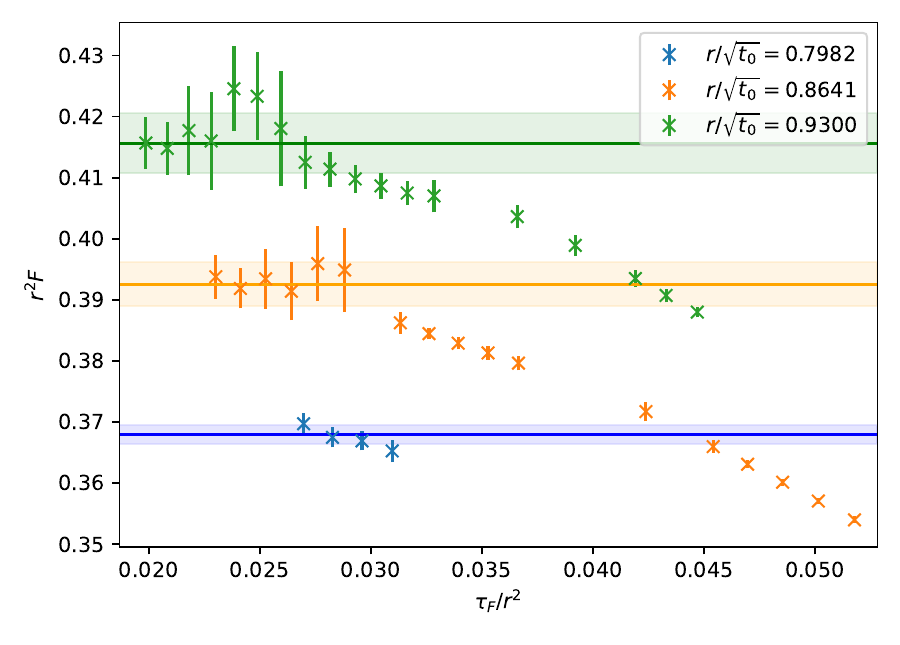}
\end{subfigure}
\begin{subfigure}{.45\textwidth}
    \centering
    \includegraphics[width=\textwidth]{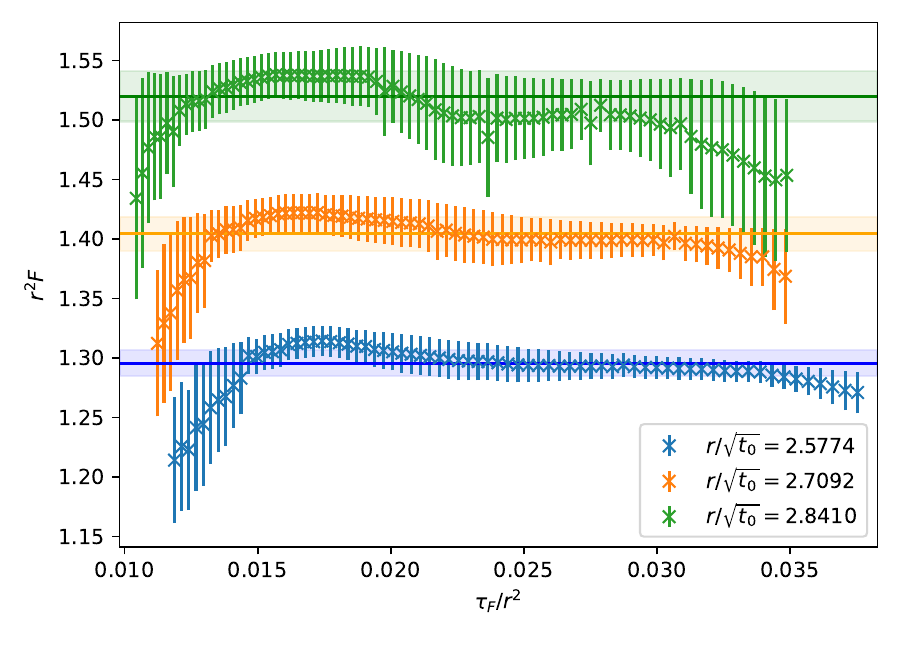}
\end{subfigure}
\caption{Examples of the flowed force in a regime where a constant behavior can be obtained. The left side shows the data for the smallest $r$ where a constant regime can be obtained. The right side shows the flowed force at larger $r$, where we can apply a constant fit in a proper range of $\tau_F/r^2$.}
\label{fig:constant_force_data}
\end{figure}

We know from Eq. \eqref{eq:NLO_force_flowed} that the static force has a constant behavior at small flow times. Physical quantities are defined at zero flow time; hence, we need to perform the zero-flow-time limit, $\tau_F\rightarrow 0$, while we keep $r$ fixed. In the constant regime, we perform this by a constant fit at fixed distance $r$. Figure~\ref{fig:constant_force_data} shows data where we obtain a constant behavior of the flowed force. The left side shows the data for the smallest $r$ before the smaller flow time comes into conflict with the $\sqrt{8\tau_F}>a$ condition. The right side shows data at larger $r$ values, where the condition $\sqrt{8\tau_F}>a$ is fulfilled at even minimal flow-time ratios. The small-flow-time expansion is performed in the ratio $\tau_F/r^2$, thus, small flow time is defined in the sense of small flow time ratio, which is a dimensionless quantity. The condition $\sqrt{8\tau_F}>a$ is given in terms of flow times in physical units; hence, considering this condition in terms of ratio  moves it to smaller flow-time ratios for larger $r$, since the $r$ in the denominator decreases the ratio.

\begin{figure}
    \centering
    \includegraphics[width=0.5\textwidth]{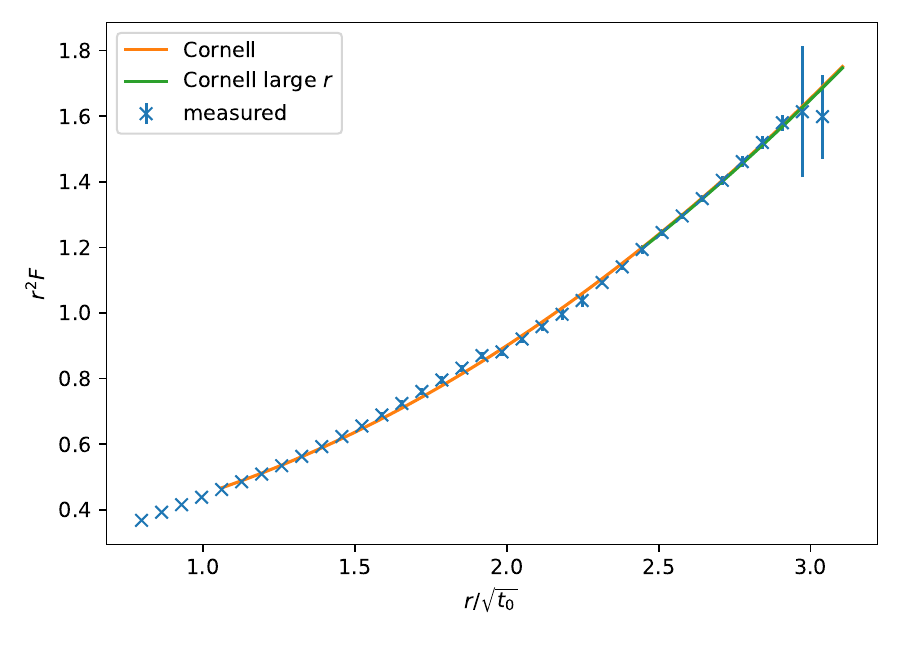}
    \caption{The extracted force from the constant zero-flow-time limits with two Cornell fits. In a window from $r/\sqrt{t_0}\approx 2.1$ to $r/\sqrt{t_0}\approx 2.3$, the continuum limit turns out to be difficult to reach. We lose the signal at larger $r$ around $r/\sqrt{t_0}\approx 3.0$.}
    \label{fig:constant_zftl_force}
\end{figure}

In Fig.~\ref{fig:constant_zftl_force}, we see the final result of the constant zero-flow-time limits. We identify an increasing behavior with small errors up to $r/\sqrt{t_0}\approx 3.0$. Around the distance of $r/\sqrt{t_0}\approx 2.25$, we are faced with difficulty extrapolation to the continuum, and we obtain $\chi^2/\mathrm{d.o.f.}$ of order 4 and larger.

We perform Cornell fit to the data from $r/\sqrt{t_0}\approx 1.1$ to $r/\sqrt{t_0}\approx 3.0$, and we obtain
\begin{align}
    r^2F(r) &= A + \sigma \ r^2 ,\\
    A &= \num{0.297(6)} \,,\\
    \sigma t_0 &= 0.151(3).
\end{align}
The string tension $\sigma$ is a quantity that is dominated by the large-$r$ regime; hence, in addition, we perform the fit only for data beyond the region where the continuum limits are problematic up to $r/\sqrt{t_0}\approx 3.0$. In this case, we obtain
\begin{align}
    A &= \num{0.268(33)}\,, \\
    \sigma t_0 &= 0.154(6).
\end{align}
The uncertainty for $A$ is 5 times larger now, which is to be expected, since it is a small $r$ quantity. The results for $\sigma$ from both ranges agree within their uncertainties.
With the result in Eq.~\eqref{eq:sqrt8t0r0_result}, we can express the string tension in $r_0$ units:
\begin{equation}
    \sigma r_0^2 = 1.345(54).
\end{equation}

\begin{figure}[ht]
    \centering
    \includegraphics[width=0.47\textwidth]{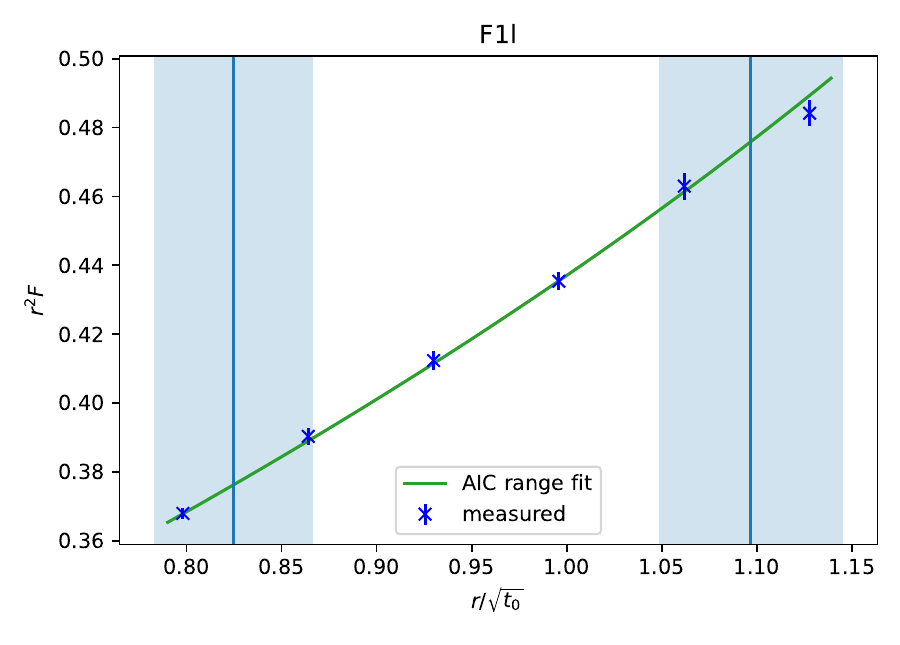}
    \includegraphics[width=0.47\textwidth]{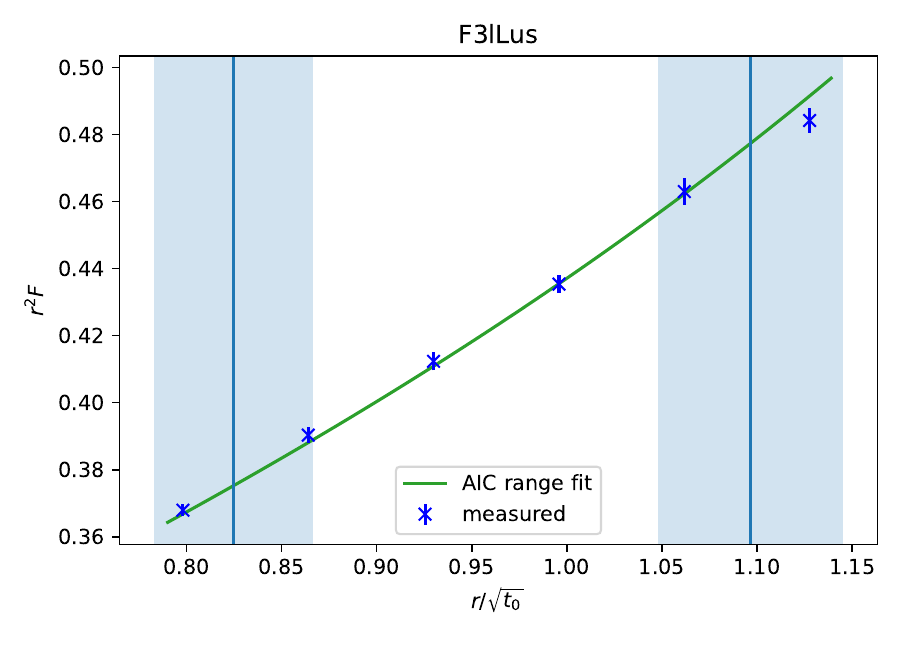}\caption{The $\sqrt{8t_0}\Lambda_0$ extraction with constant zero-flow-times extrapolation at the smallest possible distances $r$. The vertical lines with the dimmer bar represent the lower and upper averaged fit range according to Eqs. \eqref{eq:AIC_lower_fit_range} and \eqref{eq:AIC_upper_fit_range}.}
    \label{fig:lambda_with_constant_zftl}
\end{figure}

In the past, the string tension was found to be $\sigma r_0^2=1.353(3)$~\cite{Koma:2007jq} at finite lattice spacing with $\beta=6.0$, which is in good agreement with our result. Continuum results were obtained in~\cite{Gonzalez-Arroyo:2012euf,Okawa:2014kgi} in another reference scale $\Bar{r}$. With the ratio $\sqrt{8t_0}/\Bar{r}$ from~\cite{Okawa:2014kgi}, these results become $\sigma t_0=0.133(2)$ and $\sigma t_0=0.143(2)$, respectively. Nevertheless, the ratio $\sqrt{8t_0}/\Bar{r}$ in~\cite{Okawa:2014kgi} is only an approximation over several lattice sizes, and it is not extrapolated to the continuum limit. Reliable continuum results for the string tension can be found in Refs.~\cite{Lucini:2003zr,Beinlich:1997ia} in units of the critical temperature $T_c$. With the conversion factor $T_c\sqrt{t_0}$ from~\cite{Francis:2015lha}, these results become $\sigma t_0=0.1484(22)$ and $\sigma t_0=0.156(3)$, which agree well with our result.

\begin{table}[ht]
    \centering
    \begin{tabular}{c|c|c|c|c|c}
        Order & $\sqrt{8t_0}\Lambda_0$ & $\sigma_\mathrm{stat}$ & $\sigma_\mathrm{AIC}$ & $\sigma_\mathrm{stat+AIC}$\\\hline
        F1l	 & 0.8214	 & 0.0044    	 & 0.0018	 & 0.0047 \\
        F2l	 & 0.6635	 & 0.0048    	 & 0.0029	 & 0.0056\\
        F2lLus	 & 0.6961	 & 0.0057    	 & 0.0039	 & 0.0069\\
        F3l	 & 0.6197	 & 0.0036    	 & 0.0024	 & 0.0043\\
        F3lLus	 & 0.6353	 & 0.0032    	 & 0.0013	 & 0.0035\\
    \end{tabular}
    \caption{The fit result for $\sqrt{t_0}\Lambda_0$ from the constant zero-flow-time limit extracted data at different orders. 
    We see that the error is dominated by statistical fluctuations rather than the choice of the proper fit window.}
    \label{tab:final_lambda_results_fixed_czftl_force}
\end{table}

At small $r$, we extract $\Lambda_0$ from the data by fitting the perturbative force at the available orders. We solve the renormalization group equation for the running of the coupling $\alpha_\mathrm{s}$ numerically by using the \texttt{RunDec} package~\cite{Chetyrkin:2000yt,Schmidt:2012az,Herren:2017osy}, which takes $\Lambda_0$ as an input parameter. Setting the scale $\mu$ to a specific choice, we remain with a fit function depending solely on the parameter $\Lambda_0$, which serves as the fit parameter. The perturbative coefficients can be found in the literature ~\cite{Brambilla:1999qa,Pineda:2000gza,Brambilla:2006wp,Brambilla:2009bi,Anzai:2009tm,Smirnov:2009fh}, and an explicit equation for the force can be found in Eqs. (10) and (11) of~\cite{Bazavov:2014soa}.
Figure~\ref{fig:lambda_with_constant_zftl} shows examples of the fit for two different orders. Table~\ref{tab:final_lambda_results_fixed_czftl_force} shows the fit results. 
We observe that the error is dominated by the statistical fluctuations rather than by the systematic uncertainty due to the choice of the fit ranges.

\subsection{Flow-time behavior of the force at fixed $r$ or fixed $\tau_F$}\label{sec:small_r}
In the very small-$r$ regime, the requirement $\sqrt{8\tau_F}>a$ moves the data points along the 
$\tau_F/r^2$ axis outside the regime where they are flow-time independent even for the smallest possible flow time.
Therefore, we compare the lattice data with the full expression of the force, Eq.~\eqref{eq:flowed_force_given_order} -- i.e., without expanding for small $\tau_F$.

We inspect the small-$r$ flow-time behavior first at fixed distances $r$, which corresponds to the classical zero-flow-time limit. 
In a second approach, we fit along the $r$ axis at fixed flow times to
extract $\sqrt{8t_0}\Lambda_0$. Since the dependence on $\tau_F$
of the numerical extraction turns out to be negligible within the distance
and flow-time regions used for the fits to the lattice
data, getting $\sqrt{8t_0}\Lambda_0$ provides in practice its zero-flow-time
limit.

\subsubsection{Fixed $r$}\label{sec:small_fixed_r}

\begin{figure}
\centering
\begin{subfigure}{.47\textwidth}
    \centering
    \includegraphics[width=\textwidth]{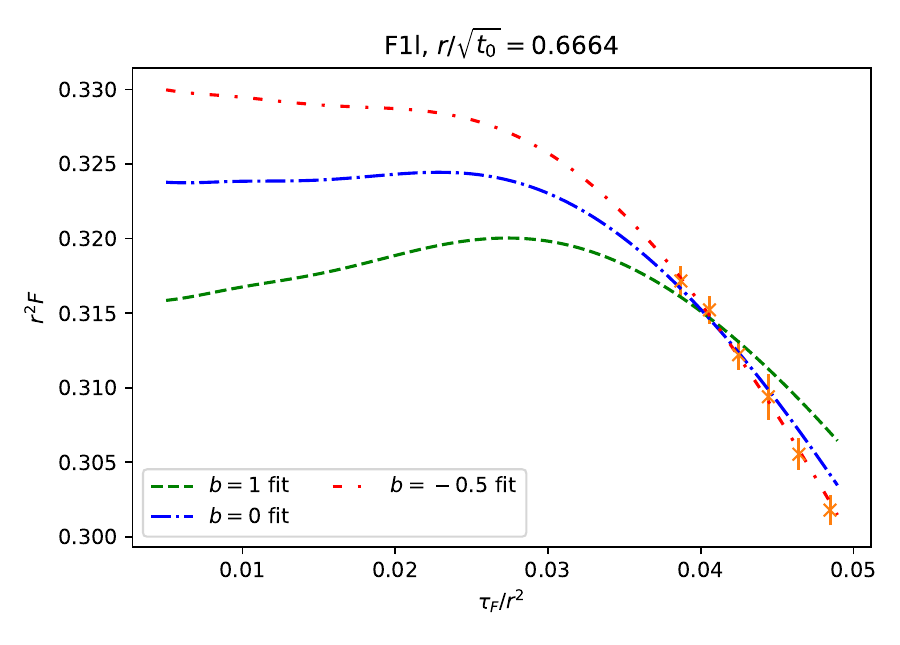}
\end{subfigure}
\begin{subfigure}{.47\textwidth}
    \centering
    \includegraphics[width=\textwidth]{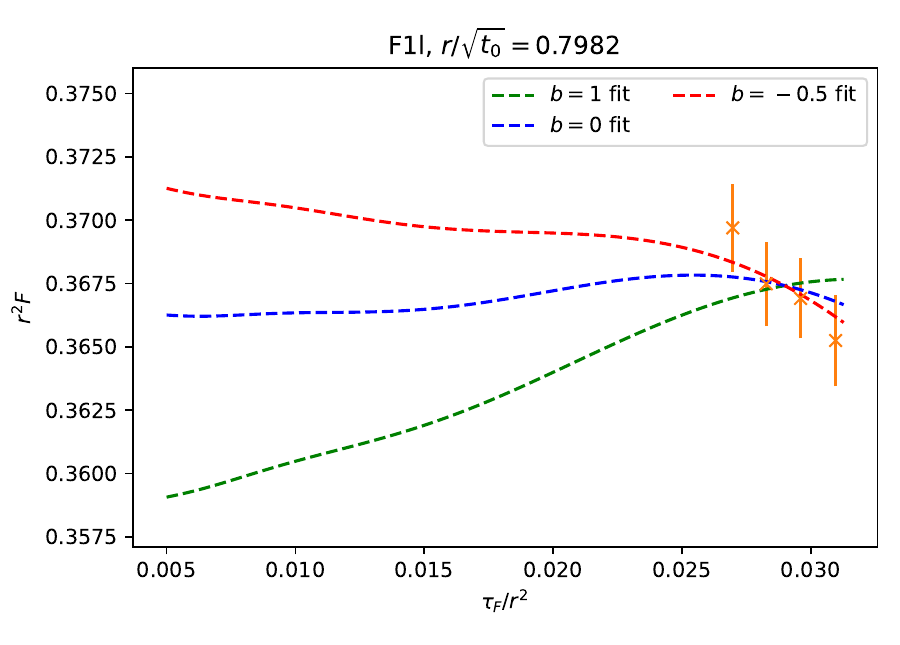}
\end{subfigure}
\caption{Comparison of different zero-flow-time limits at one-loop order at fixed $r$. We compare fits with different scale choices obtained by varying $b$ according to Eq.~\eqref{eq:generic_mu_b}.}
\label{fig:fixed_r_F1l}
\end{figure}

\begin{figure}
\centering
\begin{subfigure}{.47\textwidth}
    \centering
    \includegraphics[width=\textwidth]{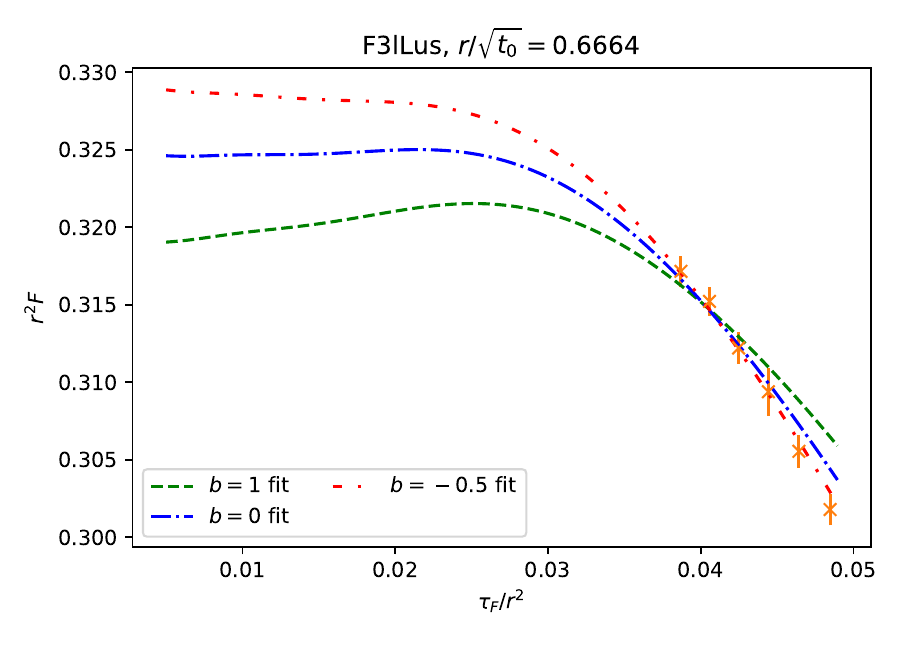}
\end{subfigure}
\begin{subfigure}{.47\textwidth}
    \centering
    \includegraphics[width=\textwidth]{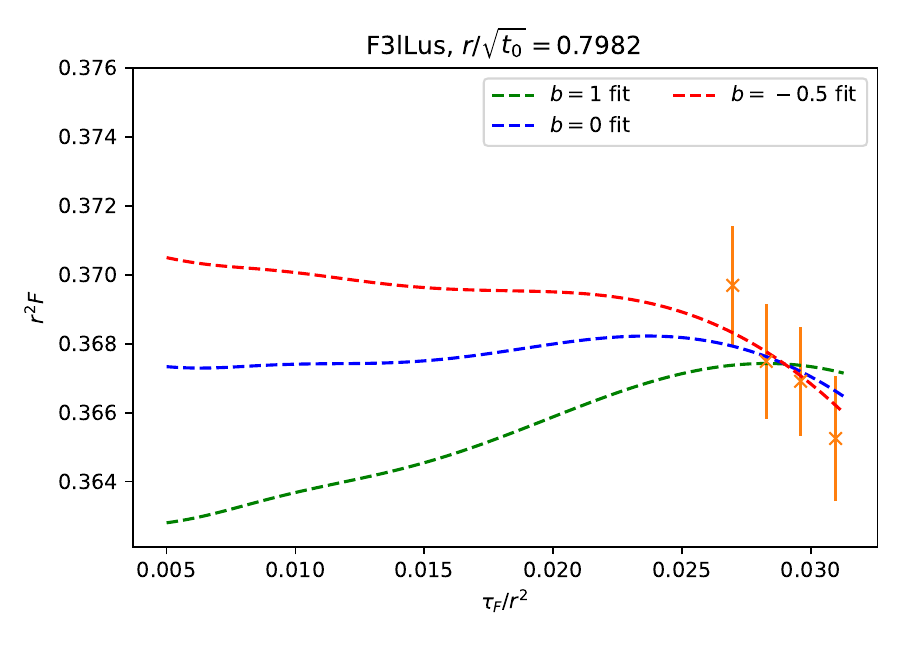}
\end{subfigure}
\caption{As in Fig. \ref{fig:fixed_r_F1l}, but with the fit function now being the force at three loops with leading ultrasoft resummation.}
\label{fig:fixed_r_F3lLus}
\end{figure}

\begin{table}[]
    \centering
    \begin{subtable}{.7\textwidth}
    \begin{tabular}{c|c|c|c|c|c|c|c}
        Order & $r/\sqrt{t_0}$ & $b$ & $\sqrt{8t_0}\Lambda_0$ & $\sigma_\mathrm{stat}$ & $\sigma_\mathrm{AIC}$ & $\chi^2/dof$ & $\overline{\tau_F/r^2}_\mathrm{Max}$\\\hline
F3lLus     & 0.6664  & 1       & 0.6020  & 0.0030  & 0.0010  & 0.57    & 0.0449(9)\\ 
           &         & 0       & 0.6229  & 0.0031  & 0.0010  & 0.52    & 0.0463(16)\\ 
           &         & -0.5    & 0.6388  & 0.0029  & 0.0005  & 0.22    & 0.0473(14)\\ 
           & 0.7323  & 1       & 0.6128  & 0.0037  & 0.0011  & 0.93    & 0.0381(14)\\ 
           &         & 0       & 0.6300  & 0.0037  & 0.0009  & 0.37    & 0.0395(20)\\ 
           &         & -0.5    & 0.6427  & 0.0037  & 0.0004  & 0.21    & 0.0403(19)\\ 
           & 0.7982  & 1       & 0.6210  & 0.0036  &                   & 0.72    &  \\ 
           &         & 0       & 0.6348  & 0.0037  &                   & 0.39    &  \\ 
           &         & -0.5    & 0.6440  & 0.0038  &                   & 0.23    &  \\ \hline
F2l        & 0.6664  & 1       & 0.6132  & 0.0032  & 0.0010  & 0.59    & 0.0449(9)\\ 
           &         & 0       & 0.6361  & 0.0033  & 0.0011  & 0.51    & 0.0463(16)\\ 
           &         & -0.5    & 0.6538  & 0.0031  & 0.0005  & 0.20    & 0.0474(14)\\ 
           & 0.7323  & 1       & 0.6274  & 0.0040  & 0.0012  & 0.97    & 0.0381(13)\\ 
           &         & 0       & 0.6468  & 0.0041  & 0.0010  & 0.38    & 0.0395(20)\\ 
           &         & -0.5    & 0.6612  & 0.0040  & 0.0005  & 0.20    & 0.0404(19)\\ 
           & 0.7982  & 1       & 0.6394  & 0.0040  &                   & 0.77    &  \\ 
           &         & 0       & 0.6555  & 0.0041  &                   & 0.41    &  \\ 
           &         & -0.5    & 0.6663  & 0.0042  &                   & 0.23    &  \\ \hline
F1l        & 0.6664  & 1       & 0.7550  & 0.0036  & 0.0010  & 0.86    & 0.0446(6)\\ 
           &         & 0       & 0.7952  & 0.0038  & 0.0012  & 0.43    & 0.0466(16)\\ 
           &         & -0.5    & 0.8276  & 0.0037  & 0.0001  & 0.04    & 0.0476(13)\\ 
           & 0.7323  & 1       & 0.7696  & 0.0045  & 0.0013  & 1.33    & 0.0376(9)\\ 
           &         & 0       & 0.8042  & 0.0048  & 0.0013  & 0.49    & 0.0393(20)\\ 
           &         & -0.5    & 0.8304  & 0.0048  & 0.0003  & 0.12    & 0.0405(19)\\ 
           & 0.7982  & 1       & 0.7829  & 0.0045  &                   & 1.29    &  \\ 
           &         & 0       & 0.8125  & 0.0048  &                   & 0.59    &  \\ 
           &         & -0.5    & 0.8328  & 0.0050  &                   & 0.26    &  \\ \hline

    \end{tabular}
    \end{subtable}
    \caption{Fit results at fixed $r$ along the flow-time axis for three different choices of the perturbative order of the static force. We start at the smallest possible flow time and use an Akaike range fit for different upper limits. No value for $\sigma_\mathrm{AIC}$ indicates that there are not enough points along the flow time to perform an Akaike range fit. The given $\chi^2/\mathrm{d.o.f.}$ corresponds to the most likely fit range. $\overline{\tau_F/r^2}_\mathrm{Max}$ gives the Akaike averaged upper flow-time limit.}
    \label{tab:fixed_r_fit_along_tauf}
\end{table}

Figures~\ref{fig:fixed_r_F1l} and ~\ref{fig:fixed_r_F3lLus} show the flow-time behavior of the force at two different fixed $r$ values along the flow-time axis. We compare our lattice data with the perturbative expressions of the force at different orders and fit them to the data. 
$\Lambda_0$ serves as the fit parameter. In the figures, we show the fit results for the case of different, but fixed choices of $b$ 
in Eq.~\eqref{eq:generic_mu_b}.
The fit range starts at the smallest possible flow time. 
For the upper bound, we take an Akaike average~\cite{Jay:2020jkz} over different fit ranges to reduce the systematics due to the fit  range choice.

From the figures, we get that the choice of the scale parameter $b$ has a definitive effect on how well the perturbative curve describes the data. A value of $b=1$ guarantees the correct asymptotic behavior at large flow time~\cite{Brambilla:2021egm}. However, we see that $b=1$ is the worst of our $b$ choices at describing the actual lattice data in the range of flow times we are interested in. 
Hence, we use negative values of $b$, which still lead to valid scaling in our range of flow times, as discussed in Appendix~\ref{sec:appendixbvalues}.
At small $r$ (left-side plots in Figs.~\ref{fig:fixed_r_F1l} and~\ref{fig:fixed_r_F3lLus}), the slope along the flow time is strong, while at the largest $r$ to which we can reasonably apply fixed-$r$ fits (right-side plots in Figs.~\ref{fig:fixed_r_F1l} and \ref{fig:fixed_r_F3lLus}), the data points seem already sensitive to the constant behavior of the force expected at small flow times. At small flow times within our considered flow-time range, the fits with $b=0$ and $b=-0.5$ agree reasonably well with the data. Table \ref{tab:fixed_r_fit_along_tauf} shows an incomplete part of the fit results.

We conclude that for smaller $r$, which requires going to larger flow-time ratios $\tau_F/r^2$, a negative value of $b$ describes the data better in the large-flow-time regime than $b=0$ or 1,  whereas  for larger $r$ all choices might describe the data within the given uncertainties. That means that fixing $b$ introduces another source of uncertainty which has to be considered. In the zero-flow-time limit, all choices of $b$ give $\mu = 1/r$, which is the natural scale choice for the static force and energy at zero flow time.

\subsubsection{Fixed $\tau_F$}\label{sec:small_r_fixed_tf}

\begin{figure}
\centering
\begin{subfigure}{.47\textwidth}
    \centering
    \includegraphics[width=\textwidth]{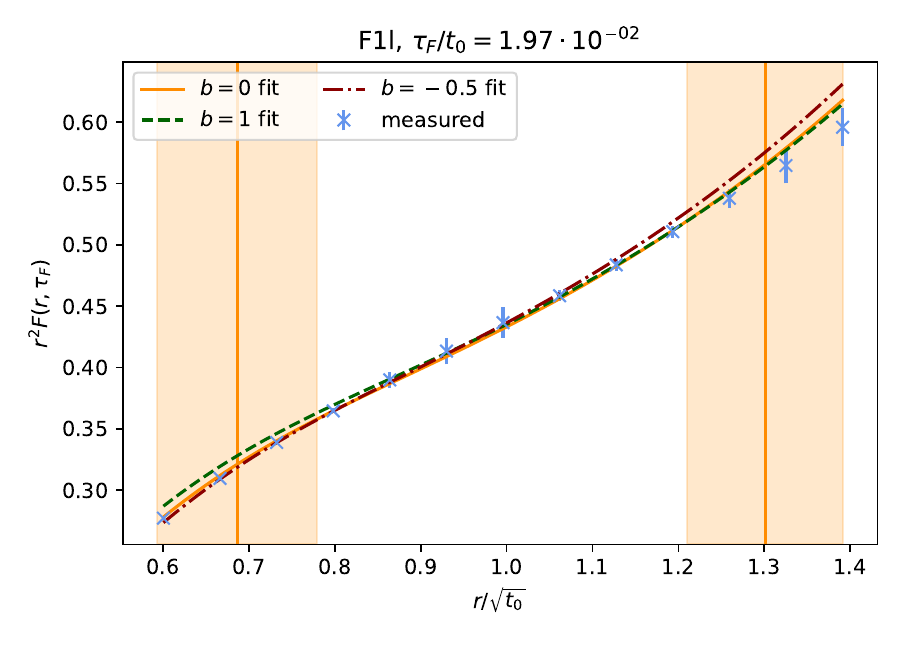}
\end{subfigure}
\begin{subfigure}{.47\textwidth}
    \centering
    \includegraphics[width=\textwidth]{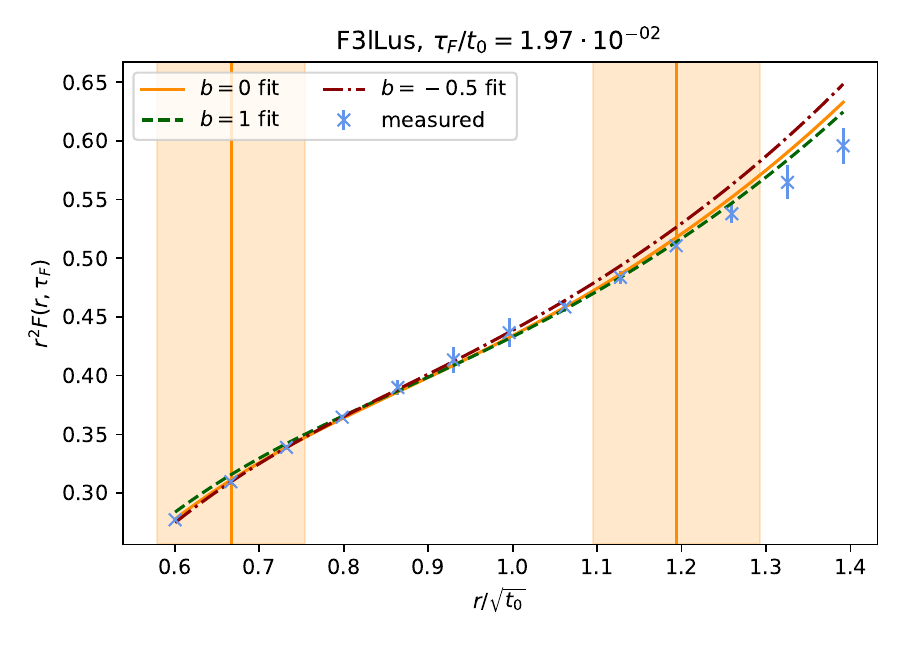}
\end{subfigure}
\caption{$\sqrt{8t_0}\Lambda_0$ fit at fixed flow time for different orders. The left vertical line corresponds to the average lower fit limit for $b=0$, and the right line to its average upper fit limit. The bands around the lines represent the errors of the fit window.}
\label{fig:fixed_t_fits_example}
\end{figure}

\begin{figure}
\centering
\begin{subfigure}{.47\textwidth}
    \centering
    \includegraphics[width=\textwidth]{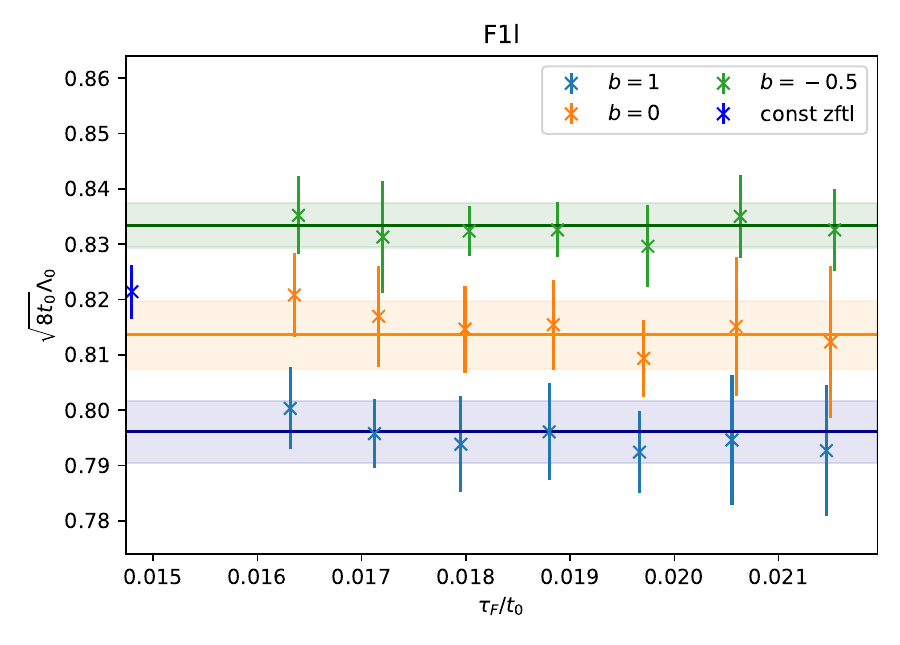}
\end{subfigure}
\begin{subfigure}{.47\textwidth}
    \centering
    \includegraphics[width=\textwidth]{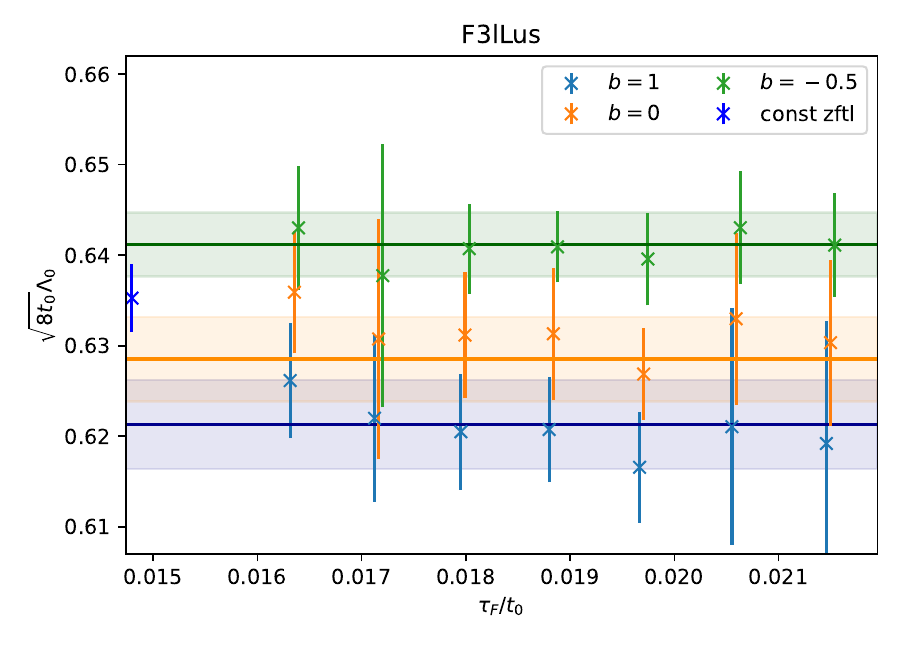}
\end{subfigure}
\caption{$\sqrt{8t_0}\Lambda_0$  in the zero-flow-time limit from a constant fit including the statistical and fit errors. We compare different orders and different choices of $b$. The blue points with the label ``const zftl'' are the results from Table \ref{tab:final_lambda_results_fixed_czftl_force}.}
\label{fig:fixed_t_lambda_fits}
\end{figure}

\begin{table}[]
    \centering
    \begin{tabular}{c|c|c|c|c|c}
        $b$-scale & F1l & F2l & F2lLus & F3l & F3lLus \\\hline
        1	 & 0.7972(56)	 & 0.6591(49)	 & 0.6911(52)	 & 0.6062(52)	 & 0.6218(50)\\
        0	 & 0.8134(57)	 & 0.6649(47)	 & 0.6982(54)	 & 0.6154(44)	 & 0.6287(44)\\
        -0.5	 & 0.8334(42)	 & 0.6709(47)	 & 0.7017(53)	 & 0.6285(35)	 & 0.6415(36)\\
    \end{tabular}
    \caption{Results of the $\sqrt{8t_0}\Lambda_0$ extraction from the constant zero-flow-time limit of $\Lambda_0$ at fixed flow time. The error includes the statistical and the AIC error from the 
    fit.}
    \label{tab:final_lambda_results_fixed_t}
\end{table}

In the next step, we use data at fixed flow times $\tau_F$ and fit the perturbative force along the $r$ axis. We use an Akaike average \cite{Jay:2020jkz} over different fit windows for reducing systematics by choosing the right fit window.
We perform one-parametric fits at fixed $b$ for $\sqrt{8t_0}\Lambda_0$.
Figure~\ref{fig:fixed_t_fits_example} shows an example fit for $b=0,1,-0.5$ for F1l and F3lLus at the same flow time. The left vertical line with the dimmer band corresponds to the average lower fit limit for the $b=0$ fit, and the right vertical line to its average upper limit.

We observe that the $b=0$ fit describes the data over a wide $r$ range from small to larger $r$. The $b=1$ fit describes the data around $r/\sqrt{t_0}=1$ and up to larger $r$ in the same way as the $b=0$ fit, but it deviates from the data at smaller $r$ in contrast to the $b=0$ fit. This matches with its lower fit range limit being at larger-$r$, but the upper fit range limit being the same as in the $b=0$ fit, indicating that the effective fit range for $b=1$ is more on the larger $r$ side. The $b=-0.5$ fit describes the data around $r/\sqrt{t_0}=1$ like the $b=0$ and $b=1$ fits, but in contrast to the $b=1$ case, it fits better to the data at small $r$ and deviates from the data at larger $r$. This matches with its upper fit range limit being at smaller $r$, but the lower fit range limit being the same as in the $b=0$ fit, indicating that the effective fit range for $b=-0.5$ is more on the smaller-$r$ side. We conclude that the range for $b$ from $-0.5$ to $1$ fits well to the data, but with different effective fit ranges. The $b=0$ fit has the widest effective fit range and will serve as our default choice.

Figure~\ref{fig:fixed_t_lambda_fits} shows the fit results for $\sqrt{8t_0}\Lambda_0$ at the valid flow-time positions for $b=1,0,-0.5$ at F1l and F3lLus. 
We observe that for a fixed choice of $b$, the values of $\Lambda_0$ are constant along the flow-time axis.
This indicates that the flow-time dependence of the static force at finite flow-time 
in the distance and flow time ranges explored in this fixed $\tau_F$ analysis 
is well captured by a constant one-loop gradient flow correction to the static force.
We then extrapolate $\Lambda_0$ to the zero-flow-time limit with a constant function.
The final results of the constant zero-flow-time limits are shown in Table~\ref{tab:final_lambda_results_fixed_t}.

\subsection{Estimate of the perturbative systematic uncertainties and final results}
\begin{figure}
\centering
\begin{subfigure}{.47\textwidth}
    \centering
    \includegraphics[width=\textwidth]{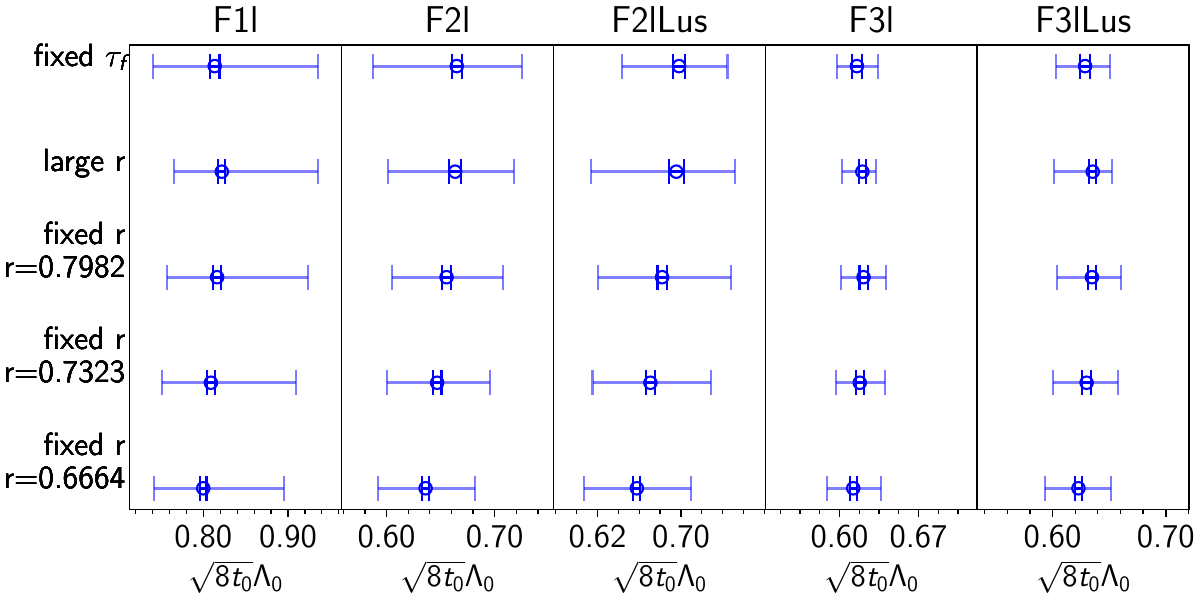}
\end{subfigure}
\begin{subfigure}{.47\textwidth}
    \centering
    \includegraphics[width=\textwidth]{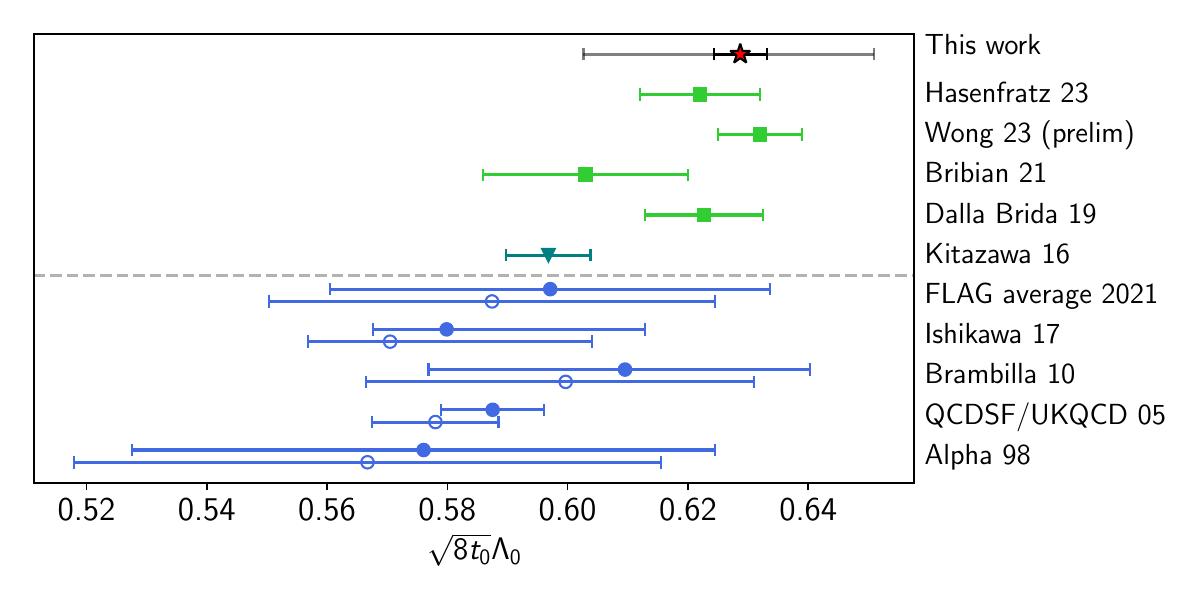}
\end{subfigure}
\caption{The final results for $\sqrt{8t_0}\Lambda_0$. Left: Comparison of all fit results for all methods and for all orders. The inner error bar corresponds to the combined statistical and AIC errors, when applicable. The outer error bar represents the total error, including the systematic uncertainties from the scale variation. 
The row ``large $r$'' shows the results from the first method, where we perform the constant zero-flow-time limit of the force first. Right: Our final result compared to the literature. The points above the dashed line were determined with gradient-flow-based scale settings, $t_0$ (squares) or $w_0$ (triangle), while the lower points (circles) are converted from $r_0$ units with either our ratio in Eq.~\eqref{eq:sqrt8t0r0_result} (filled points), or with the ratio from~\cite{DallaBrida:2019wur} (hollow points).}
\label{fig:final_results}
\end{figure}

Up to this point, we have presented results for $\Lambda_0$ with error estimates 
that include only the statistical errors
and the systematic errors from choosing different fit ranges. We still need to include the perturbative uncertainty from the unknown higher-order terms in the perturbative expansion. We can do this by varying the scale $\mu$~\eqref{eq:generic_mu_b}. In previous studies of the static energy~\cite{Bazavov:2019qoo,Bazavov:2014soa,Bazavov:2012ka}, the zero-flow-time scale $\mu=1/r$ was 
varied by a factor of $\sqrt{2}$. We make here the same choice and vary the parameter $s$ in Eq.~\eqref{eq:generic_mu_b} from $s=0.5$ to $s=2$. 
We vary the $s$ parameter only in the zero-flow-time part of  Eq.~\eqref{eq:flowed_force_given_order} and keep it fixed at $s=1$ in the  finite-flow-time part.
In principle, we could vary the scale by a factor of $2$ instead of $\sqrt{2}$, but it was noted in Ref.~\cite{Bazavov:2019qoo} that this requires access to quite  small distances $r$. Our current data do not contain  
small enough distances to allow for this wider variation. 

As already stated in the previous sections, the finite-flow-time part of the static force has a considerable dependence on the choice of the parameter $b$ in Eq.~\eqref{eq:generic_mu_b}. 
To match the conventions of zero-flow-time studies, we choose $b=0$ as our main result. 
To estimate the systematic error due to the choice of $b$ and the missing higher-order finite-flow-time perturbative terms, we vary the $b$ parameter between $b=-0.5$ and $b=1$; for this choice, we refer to the discussion in  appendix~\ref{sec:appendixbvalues}.
We vary $b$ only in the finite-flow-time part of Eq.~\eqref{eq:flowed_force_given_order}.

To arrive at a final result for $\Lambda_0$, we have explored several possibilities in Secs.~\ref{sec:large_r_constant_zftl}, \ref{sec:small_fixed_r}, and~\ref{sec:small_r_fixed_tf}:
\begin{enumerate}
\item We have performed the constant zero-flow-time limit of the force first, followed by fitting the perturbative expression to the data. This method has the advantage that we 
do not need to combine the zero-flow-time expression of the force with the one-loop correction coming from the gradient flow.
The method does not work, however, at the smallest $r$, but only for $r/\sqrt{t_0}\geq 0.8$.

\item We have performed the fit of the combined equation at fixed $r$ along $\tau_F$. This method corresponds to the classical zero-flow-time limit.  With this method, the $1/r$ scale has a minor effect, and the method can be applied for only a few $r$, while the gradient flow scale has a dominant role, which can be seen by the large dependence on the choice of $b$. 

\item We have performed the fit of the combined equation at fixed $\tau_F$ along $r$. In this method, the $1/r$ scale has a major impact, while the flow-time scale has a minor role. This can be seen by the fact that various choices of $b$ fit well to the data.
\end{enumerate}
On the left side of Fig.~\ref{fig:final_results}, we compare the results of all three methods at the available perturbative orders. 
All results agree very well within the errors.

Based on the advantages and disadvantages of all three methods, we chose method 3 with $b=0$ at F3lLus as the main result. 
Including the uncertainties due to variations of the parameters $s$ and $b$,  
we obtain our final result, which reads 
\begin{align}
    \sqrt{8t_0}\Lambda_0 &= 0.629^{+22}_{-26}\,,\\
    \delta(\sqrt{8t_0}\Lambda_0) &= (4)^\mathrm{lattice}(^{+18}_{-25})^\mathrm{s\mhyphen scale}(^{+13}_{-7})^\mathrm{b\mhyphen scale},
\end{align}
where $(...)^\mathrm{lattice}$ stands for the error coming from the statistics and choosing different fit windows. 
As we allow the parameters $b$ and $s$ in~Eq.~\eqref{eq:generic_mu_b} to be varied
only in the finite- and zero-flow-time part respectively, it follows that the systematic uncertainties from
these variations are nearly independent one from the other. Hereby, we quote the uncertainty from the $b$-scale variation measured at $s=1$ and the uncertainty from the $s$-scale variation measured at $b=0$, and we add these in quadrature. This accounts for a conservative estimate of the perturbative error.

On the right side of Fig.~\ref{fig:final_results}, we compare our final result with results from previous
measurements of $\Lambda_0$~\cite{Capitani:1998mq,Gockeler:2005rv,Brambilla:2010pp,Ishikawa:2017xam,Kitazawa:2016dsl,DallaBrida:2019wur,Bribian:2021cmg,Wong:2023jvr,Hasenfratz:2023bok} and the FLAG average~\cite{FlavourLatticeAveragingGroupFLAG:2021npn}.
We only show previous studies that contribute to the FLAG average and a couple of newer studies that have come out since the latest FLAG average. In  Fig.~\ref{fig:final_results}, the points above the dashed line have been obtained  using the gradient flow-based scales $t_0$ and $w_0$,\footnote{The measurement of Ref.~\cite{Kitazawa:2016dsl} in $w_0$ units was transformed to $t_0$ units using the ratios from~\cite{Asakawa:2015vta}.}
while the points below the dashed line 
have been obtained from the scale $r_0$. For the measurements done in the $r_0$ scale, we convert to the $t_0$ scale
with our own ratio~\eqref{eq:sqrt8t0r0_result} (filled points) or with one of the ratios from Ref.~\cite{DallaBrida:2019wur} (empty points). We find it remarkable that all the newer studies done in $t_0$ units are on the higher end of the  measurements. Furthermore, we note that our error is larger than other recent studies. Our error is dominated by the perturbative error from the scale variation. Since the scale variation is more prominent at larger distances, it is to be expected that future access to finer lattices could bring this error down.
Lastly, with the ratio in Eq.~\eqref{eq:sqrt8t0r0_result}, we can convert our final result into $r_0$ units and obtain
\begin{align}
    r_0\Lambda_0 = 0.657^{+23}_{-28}\,.
\end{align}

\section{Summary and conclusion}
We have shown that the gradient flow renormalizes an operator made of a Wilson loop with a chromoelectric field insertion by reducing discretization effects, 
and in this way improving the convergence towards the 
continuum limit. This result can be of use for further studies on operators with different field insertions, which typically show up in nonrelativistic effective field theories.

Thanks to the above property, we are able to perform the continuum limit of the static force at finite flow time, and extrapolate to zero flow time in three different ways. 
In the first method, we extrapolate the static force from a constant zero-flow-time limit. This works for large and intermediate $r$, 
but not at very short distances in the regime $r/\sqrt{t_0}<0.8$. 
At large distances, we extract the scales $r_0$ and $r_1$, and are able to perform a Cornell fit. For the scales, we find the ratios
\begin{align}
    \frac{r_0}{r_1} &= \num{1.380(14)} \,,\\
    \frac{\sqrt{8t_0}}{r_0} &= \num{0.9569(66)}\,,\label{eq:summary_t0r0_ratio}\\
    \frac{\sqrt{8t_0}}{r_1} &= \num{1.325(13)}\,,\label{eq:summary_t0r1_ratio}
\end{align}
and for the string tension parameter in the Cornell fit, we find
\begin{align}
    \sigma t_0 &= 0.154(6) \,,\\
    \sigma r_0^2 &= 1.345(54)\,,
\end{align}
where we have used our result for the ratio $\sqrt{8t_0}/r_0$ in Eq.~\eqref{eq:summary_t0r0_ratio} to convert into $r_0$ units. 
At short distances, we fit the perturbative force to the data, and we obtain at F3lLus order
\begin{equation}
    \sqrt{8t_0}\Lambda_0 = \num{0.635(4)}\,.
\end{equation}

In the second and third methods, we fit with a function that combines the force at zero flow time up to three loops with the one-loop flow time correction. In this way, the fit function depends on two scales, $r$ and $\tau_F$. In the second method, we keep $r$ fixed and perform the fit along $\tau_F$. In the third method, we keep $\tau_F$ fixed and perform the fit along $r$, and extrapolate the resulting $\Lambda_0(\tau_F)$ to the zero-flow-time limit. The third method has, in comparison to the second method, a strong dependence on the scale $1/r$, which is the dominant physical scale. Furthermore, the third method reaches out to small $r$ in contrast to the first method. Therefore, we take the third method as our  reference method and obtain
\begin{align}
    \sqrt{8t_0}\Lambda_0 &= 0.629_{-26}^{+22} \,,\\
    \delta(\sqrt{8t_0}\Lambda_0) &= (4)^\mathrm{lattice}(_{-25}^{+18})^\mathrm{s\mhyphen scale}
    (_{-7}^{+13})^\mathrm{b\mhyphen scale}.
\end{align}
Using the ratio in Eq. \eqref{eq:summary_t0r0_ratio}, we can give our final result in $r_0$ units as 
\begin{align}
    r_0\Lambda_0 = 0.657^{+23}_{-28}\,.
\end{align}
Nevertheless, all methods agree well within their errors, with an overlap of almost \SI{70}{\percent}.

\acknowledgements
We would like to thank Johannes H. Weber for useful discussions. The lattice QCD calculations were performed using the publicly available \href{https://web.physics.utah.edu/~detar/milc/milcv7.html}{MILC code}.
The simulations were carried out on the computing facilities of the Computational Center for Particle and Astrophysics (C2PAP), the Leibniz-Rechenzentrum (LRZ) in the project ``Calculation of finite-$T$ QCD correlators" (pr83pu), and the SuperMUC cluster at the LRZ in the project ``The role of the charm-quark for the QCD coupling constant" (pn56bo).
J.~M.-S. acknowledges support by the Munich Data Science Institute (MDSI) at the Technical University of Munich (TUM) via the Linde/MDSI Doctoral Fellowship program.
This research was funded by the Deutsche Forschungsgemeinschaft (DFG, German Research Foundation) cluster of excellence “ORIGINS” (\href{www.origins-cluster.de}{www.origins-cluster.de}) under Germany’s Excellence Strategy No. EXC-2094-390783311.
The work of N. B. and J. M.-S. is supported by the DFG  (Deutsche Forschungsgemeinschaft, German Research Foundation) under Grant No. BR 4058/2-2.
N. B., J. M.-S., and A. V. acknowledge support from the STRONG-2020 European Union’s Horizon 2020 research and innovation program under Grant Agreement No. 824093.

\appendix
\section{About negative values of $b$ in $\mu$}\label{sec:appendixbvalues}

\begin{figure}
    \centering
    \includegraphics[width=0.45\textwidth]{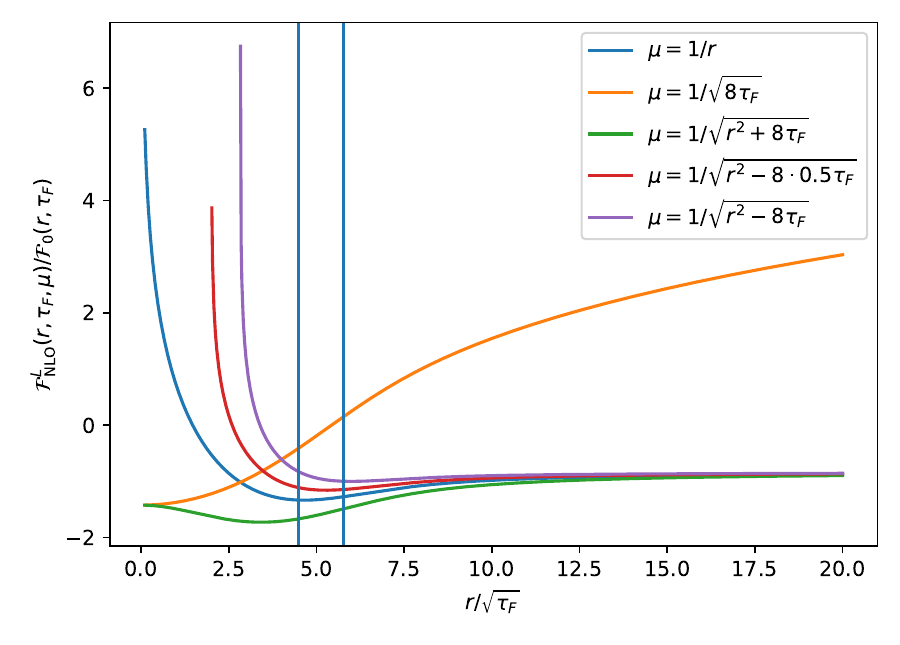}
    \includegraphics[width=0.45\textwidth]{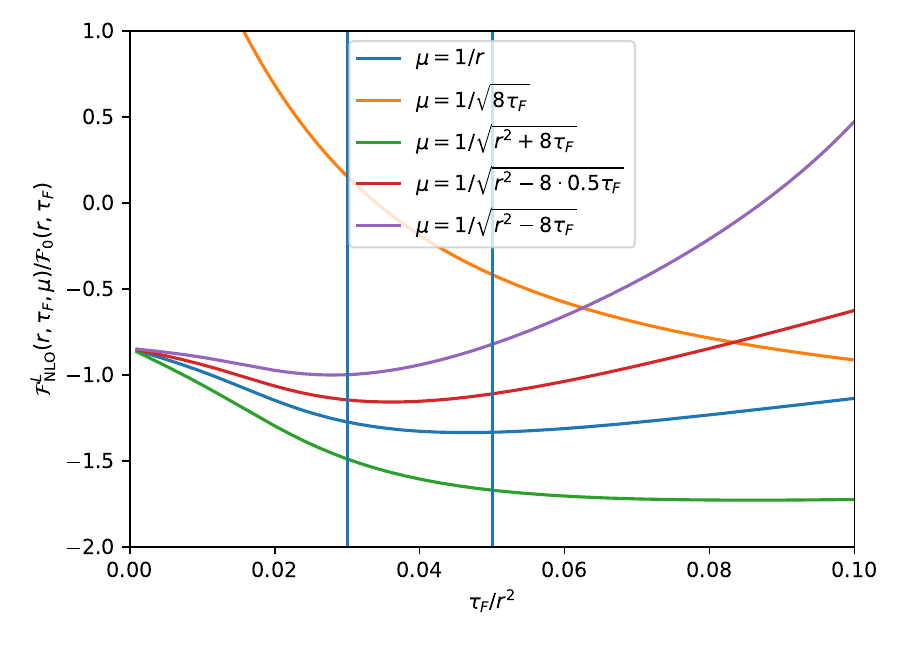}
    \caption{
    Left: The ratio $\mathcal{F}_\mathrm{NLO}^L(r,\tau_F,\mu)/\mathcal{F}_0(r,\tau_F)$
    [see Eq. \eqref{eq:flowed_force_full_1_loop_formula}] as a function of $r/\sqrt{\tau_F}$
    plotted  for various values of $s$ and $b$. The vertical lines represent the characteristic flow-time window of this study. The left limit on the $x$ axis corresponds to the infinite-flow-time limit, while the right limit corresponds to the zero-flow-time limit. 
    Right: The same ratio, but as a function of $\tau_F/r^2$ zoomed around the region of interest.}
    \label{fig:log_NLO_contribution}
\end{figure}

In the analysis done in Sec. \ref{sec:continuum_analysis}, we chose the scale \eqref{eq:generic_mu_b} with different values for $b$, including negative values. 
The ratio $\mathcal{F}_\mathrm{NLO}^L(r,\tau_F,\mu)/\mathcal{F}_0(r,\tau_F)$ for different 
values of the parameters $s$ and $b$ is shown in  Fig.~\ref{fig:log_NLO_contribution}.
In~\cite{Brambilla:2021egm}, 
the choices $b=1,0$ with $s=1$, and $b=1$ with $s=0$ were also analyzed. 
The choice $s=1$, $b=0$ is the natural choice at zero flow time, since for it, the 
$\log(\mu r)$ terms vanish. However, this choice does not capture $\log(\mu\sqrt{\tau_F})$ terms
that become important at large flow time. This is shown by the ratio $\mathcal{F}_\mathrm{NLO}^L(r,\tau_F,\mu)/\mathcal{F}_0(r,\tau_F)$ becoming large at large flow times for this choice of parameters.
The choice $s=0$, $b=1$ is the natural choice at large flow time, since for it 
$\log(\mu \sqrt{\tau_F})$ terms vanish.  However, this choice does not capture $\log(\mu r)$ terms
that become important at small flow times.
This is shown by the ratio $\mathcal{F}_\mathrm{NLO}^L(r,\tau_F,\mu)/\mathcal{F}_0(r,\tau_F)$ becoming large at small flow times 
for this choice of parameters.
The choice of $s=1$ and $b=1$ interpolates between these two extreme and provides a small 
$\mathcal{F}_\mathrm{NLO}^L(r,\tau_F,\mu)$ correction with respect to the leading gradient flow term $\mathcal{F}_0(r,\tau_F)$ over the whole range of flow times.
Also, the overall scale dependence turns out to be weak with this scale choice.

Nevertheless, our lattice data explore a very specific and limited region of flow-time values, 
the region in between the vertical lines in Fig.~\ref{fig:log_NLO_contribution}. A zoomed-in view of this region is shown in the right plot that makes manifest that different choices of $b$, keeping $s=1$, provide, indeed, even smaller and more stable corrections $\mathcal{F}_\mathrm{NLO}^L(r,\tau_F,\mu)$ in the region of interest than the choice $b=1$. In particular, this is the case for $b=0$ and $b=-0.5$, which indeed best fit 
our lattice data, as we have discussed in the main body of the paper.
More negative values of $b$ further reduce the relative size of $\mathcal{F}_\mathrm{NLO}^L(r,\tau_F,\mu)$, but they make it more scale dependent.
Hence, the one-loop expression of the gradient flow expression of the force suggests that 
for $0.03 \lesssim \tau_F/r^2 \lesssim 0.05$, the 
ideal choice of the parameter $b$ is in between 0 and a negative number larger than $-1$.
This is confirmed by the lattice data.
Clearly, also, a parametrization with negative $b$ must smoothly go over $\mu = 1/\sqrt{\tau_F}$
at large flow time. However, the specific form of the parametrization at large flow times, 
$\tau_F/r^2 > 0.05$, cannot be explored with the present data.

\section{Flow-time dependence of the Wilson loops with and without chromoelectric field insertions}\label{sec:flowed_wilson_loops}
\begin{figure}
\centering
\begin{subfigure}{.44\textwidth}
    \centering
    \includegraphics[width=0.9\textwidth]{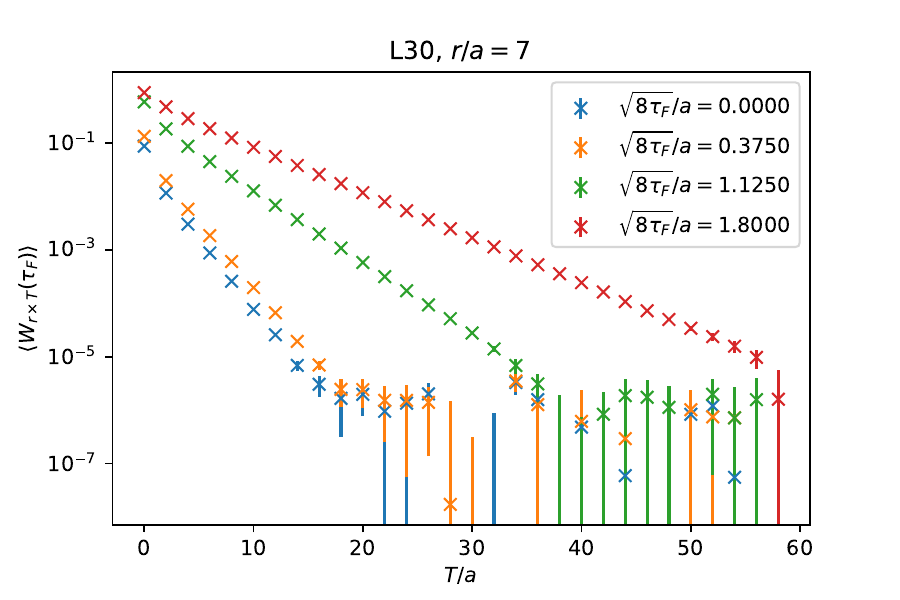}
\end{subfigure}
\begin{subfigure}{.44\textwidth}
    \centering
    \includegraphics[width=0.9\textwidth]{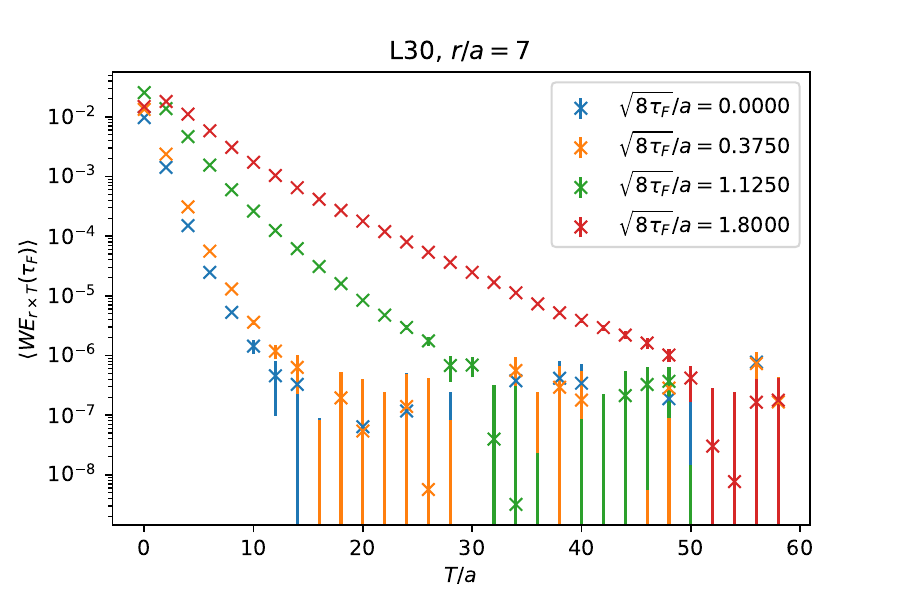}
\end{subfigure}
\begin{subfigure}{.44\textwidth}
    \centering
    \includegraphics[width=0.9\textwidth]{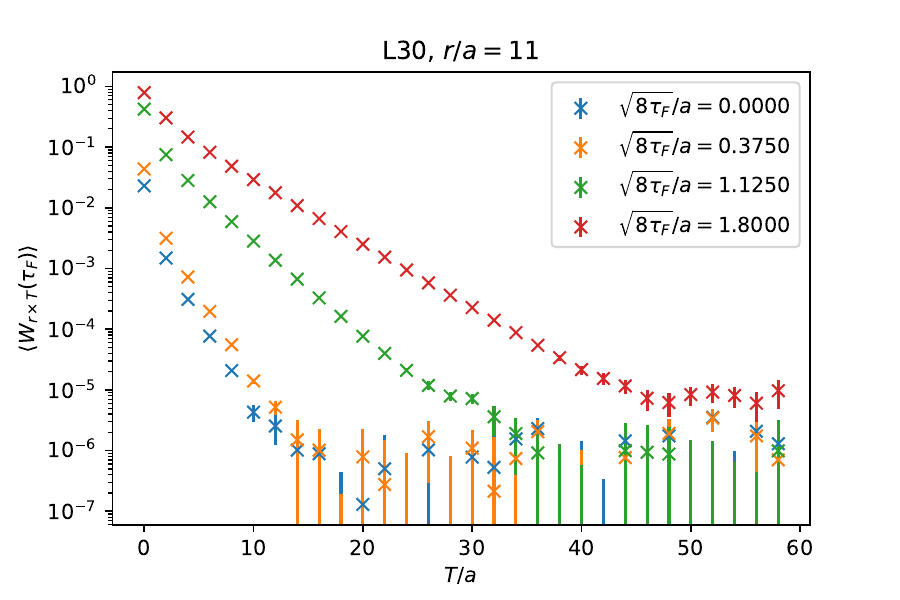}
\end{subfigure}
\begin{subfigure}{.44\textwidth}
    \centering
    \includegraphics[width=0.9\textwidth]{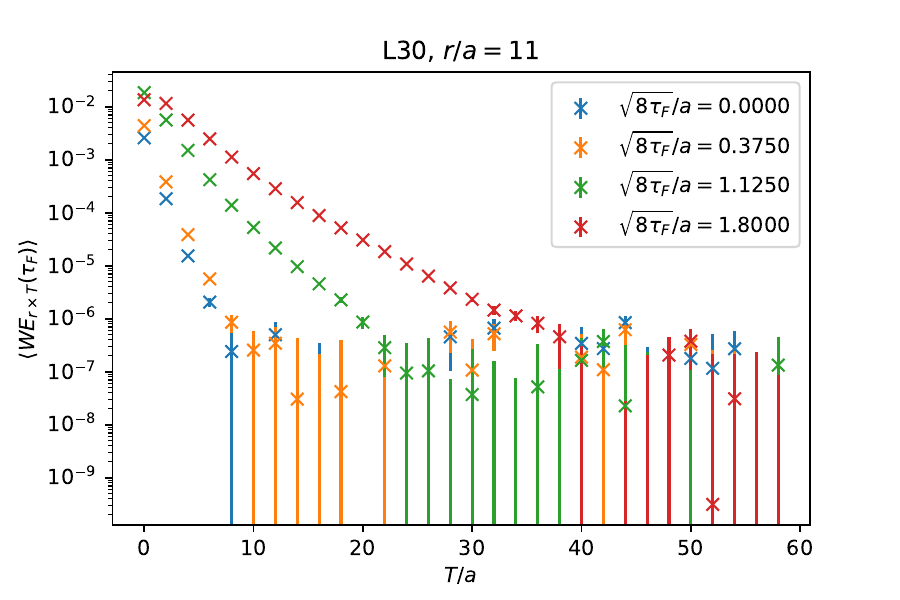}
\end{subfigure}
\caption{The left sides show the $T$ dependence of the Wilson loops at four different fixed flow times and for two fixed $r$. The right sides show the same for the Wilson loops with a chromoelectric field insertion.}
\label{fig:wilson_loops}
\end{figure}

\begin{figure}
\centering
\begin{subfigure}{.44\textwidth}
    \centering
    \includegraphics[width=0.9\textwidth]{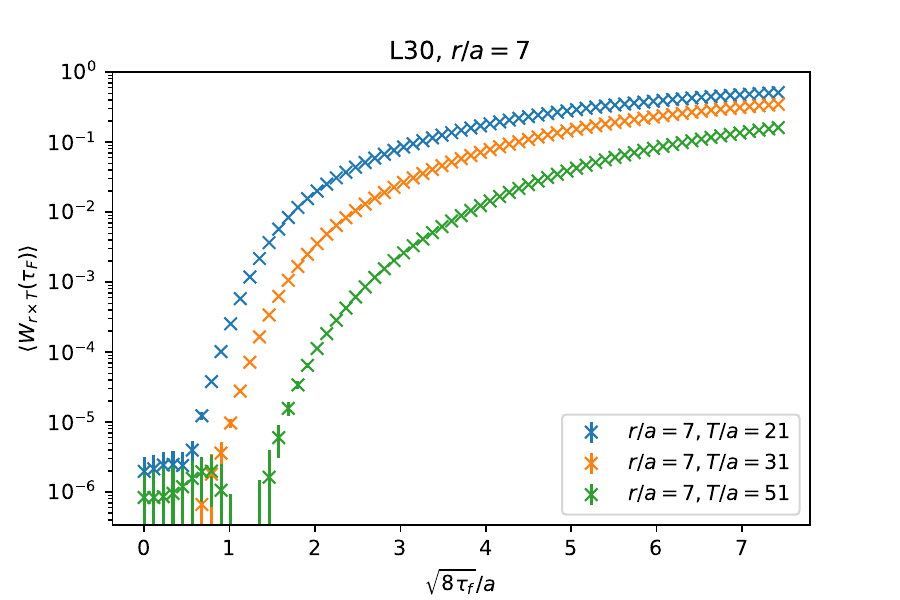}
\end{subfigure}
\begin{subfigure}{.44\textwidth}
    \centering
    \includegraphics[width=0.9\textwidth]{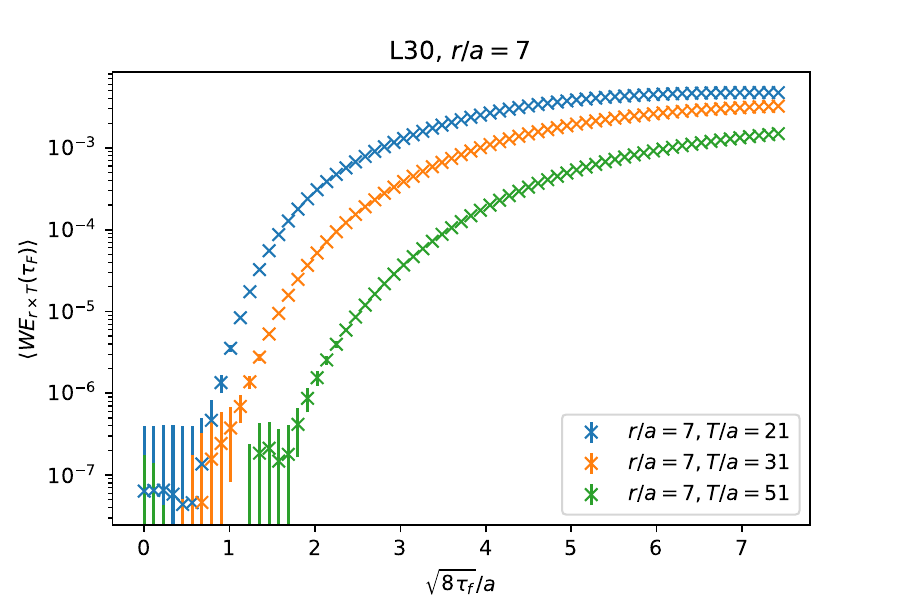}
\end{subfigure}
\begin{subfigure}{.44\textwidth}
    \centering
    \includegraphics[width=0.9\textwidth]{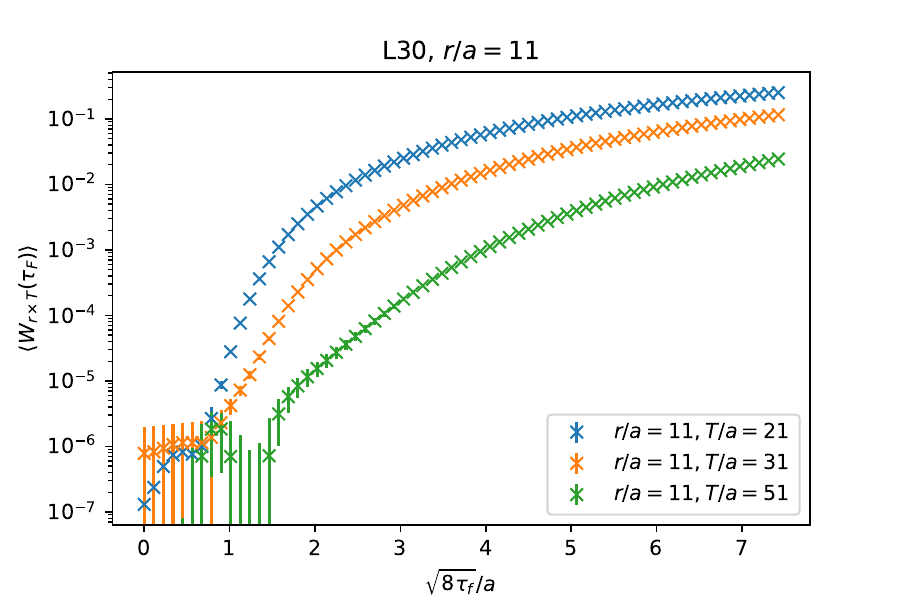}
\end{subfigure}
\begin{subfigure}{.44\textwidth}
    \centering
    \includegraphics[width=0.9\textwidth]{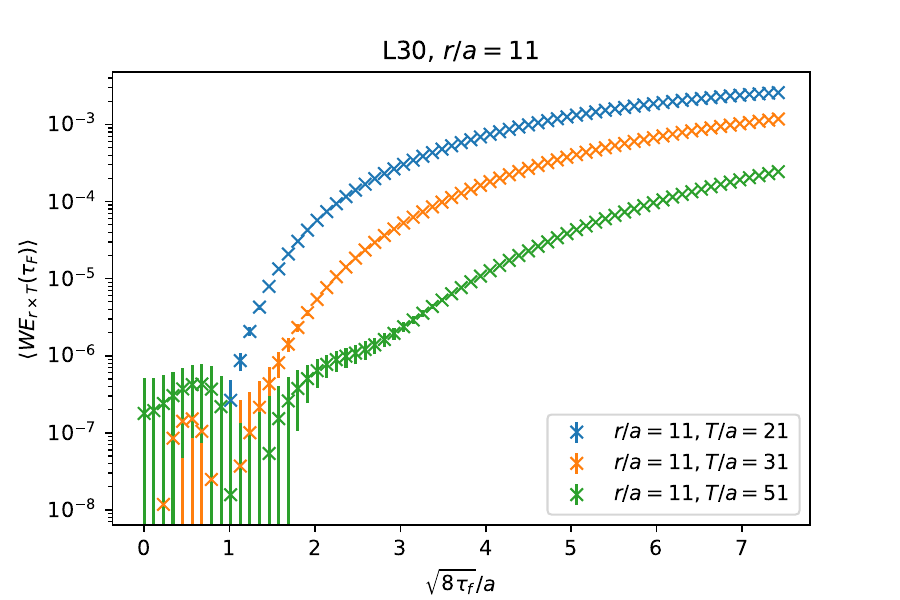}
\end{subfigure}
\caption{The left sides show the flow-time dependence of the Wilson loop at several fixed $T$ and at two different $r$ values. The right sides show the same for the Wilson loops with a chromoelectric field insertion.}
\label{fig:wilson_loops_fixedT}
\end{figure}

The Wilson loops with and without chromoelectric field insertion are the main objects of this work; therefore, it is worth to have a closer look on the flow-time dependence of them. Figure~\ref{fig:wilson_loops} shows the $T$ dependence of the Wilson loops at different flow times. In a logarithmic $y$ scale, we see linear, decreasing curves for large $T$, which correspond to the exponential falloff controlled  by the static energy. For zero and small flow times, the slope is the same; for larger flow times, the slope becomes more flat. This reflects the flow-time dependency of the static energy. This observation holds for Wilson loops with and without chromoelectric field insertions.

Figure~\ref{fig:wilson_loops_fixedT} shows the flow-time dependence of the Wilson loops  with and without chromoelectric field insertion at fixed $r$ and $T$. We see a strong flow-time dependence for both cases caused by the divergence of the static quark propagator. The strong flow-time dependence cancels in the ratio of the Wilson loops.

\section{Autocorrelation times on L40}
\label{app:autocorrelation}
\begin{figure}
    \centering
    \includegraphics[width=0.5\textwidth]{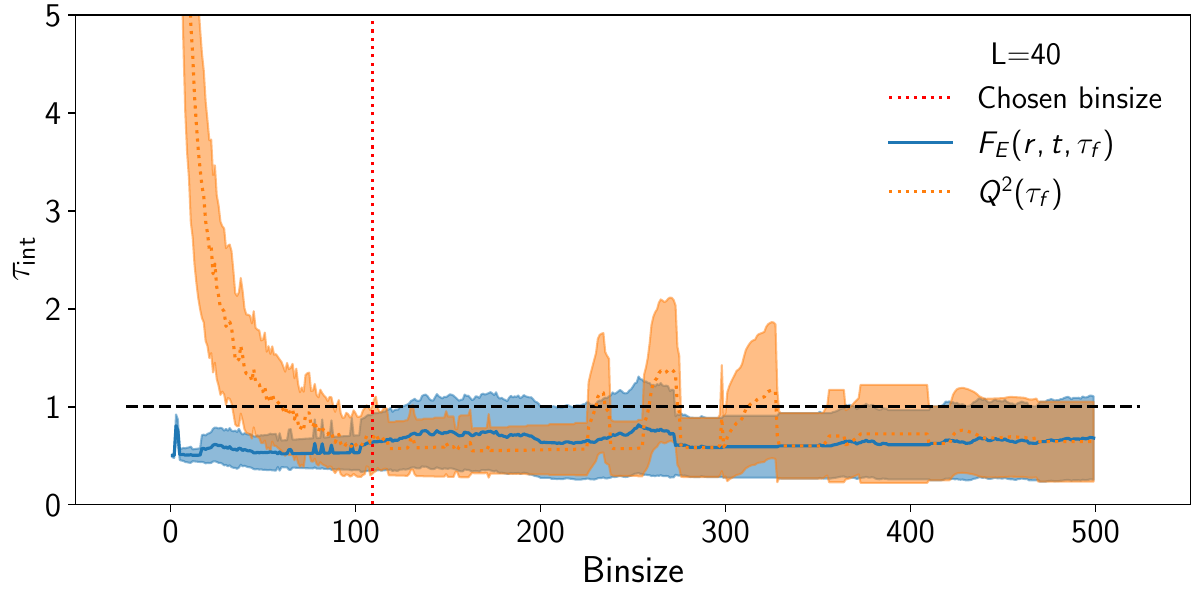}
    \caption{Integrated autocorrelation time $\tau_\mathrm{int}$ as a function of the block size for the largest lattice ensemble L40. The curve with a solid line shows the autocorrelation time for the static force, while the dashed curve shows $\tau_\mathrm{int}$ for the topological charge squared, and the vertical line indicates our chosen blocking.}
    \label{fig:tauintbinsize}
\end{figure}
To reduce the integrated autocorrelation times, we block the data to 30 jackknife blocks per ensemble. Apart from our largest lattice (L40), this choice leads to 200 configs per block, which is 
considerably higher than the integrated autocorrelation times $\tau_\mathrm{int}\lesssim\mathcal{O}(10)$ for all the relevant observables including the topological charge $Q$. 
However, we have less statistics for the L40 lattice, which leads to only 110 configurations per block. Moreover, we observe slower fluctuation of topological charge for the L40 lattice with $\tau_\mathrm{int}=101.6$ for the unblocked data. To see whether the block size is large enough also for this ensemble, we plot $\tau_\mathrm{int}(Q^2)$\footnote{We note that for the topological charge squared $Q^2$, which is related to the topological susceptibility, the integrated autocorrelation times for the unblocked data are about half of the $\tau_\mathrm{int}$ for the $Q$.}
as a function of the block size in Fig.~\ref{fig:tauintbinsize}. We observe that even the topological charge has a $\tau_\mathrm{int}$ of order 1, at our chosen block size. More importantly, we also note that the static force has autocorrelation times below 1 for all possible blocksizes. This is expected, as the static energy is known to be moderately unaffected by the topological slowing down~\cite{Weber:2018bam}. Similar topology independence has also been observed for other methods of determining the strong coupling constant~\cite{Bonanno:2024nba}.

\FloatBarrier
\bibliography{force}{}

\end{document}